\documentclass[prc,aps,floats,floatfix,showpacs,nofootinbib,superscriptaddress]{revtex4}

\usepackage{amsmath}
\usepackage{amssymb}
\usepackage{amsfonts}
\usepackage{graphicx}
\usepackage{pstricks}
\usepackage{tikz}
\usepackage{mathtools}

\newcommand{\vnabla}{\boldsymbol{\mathbf\nabla}}

\newcommand{\vPi}{\boldsymbol{\mathbf\Pi}}

\newcommand{\bbM}{\boldsymbol{\mathbb M}}
\newcommand{\bj}{\boldsymbol{\mathbf j}}

\newcommand{\br}{\boldsymbol{\mathbf r}}
\newcommand{\bs}{\boldsymbol{\mathbf s}}
\newcommand{\bJ}{\boldsymbol{\mathbf J}}
\newcommand{\bS}{\boldsymbol{\mathbf S}}
\newcommand{\bT}{\boldsymbol{\mathbf T}}

\newcommand{\nn}{\nonumber}
\newcommand{\q}{\quad}
\newcommand{\be}{\begin{equation}}
\newcommand{\ee}{\end{equation}}
\newcommand{\bqr}{\begin{eqnarray}}
\newcommand{\eqr}{\end{eqnarray}}

\begin{document}

\title{Solution of Hartree-Fock-Bogoliubov equations and fitting procedure using N2LO Skyrme pseudo-potential in spherical symmetry.}


\author{P. Becker}
\email{pbecker@ipnl.in2p3.fr}
\affiliation{Universit\'e de Lyon, Universit\'e Lyon 1,
             43 Bd. du 11 Novembre 1918, F-69622 Villeurbanne cedex, France\\
             CNRS-IN2P3, UMR 5822, Institut de Physique Nucl{\'e}aire de Lyon}
             
\author{D. Davesne}
\email{davesne@ipnl.in2p3.fr}
\affiliation{Universit\'e de Lyon, Universit\'e Lyon 1,
             43 Bd. du 11 Novembre 1918, F-69622 Villeurbanne cedex, France\\
             CNRS-IN2P3, UMR 5822, Institut de Physique Nucl{\'e}aire de Lyon}

\author{J. Meyer}
\affiliation{Universit\'e de Lyon, Universit\'e Lyon 1,
             43 Bd. du 11 Novembre 1918, F-69622 Villeurbanne cedex, France\\
             CNRS-IN2P3, UMR 5822, Institut de Physique Nucl{\'e}aire de Lyon}
             
\author{J. Navarro}
\email{navarro@ific.uv.es}
\affiliation{IFIC (CSIC-Universidad de Valencia), Apartado Postal 22085, E-46.071-Valencia, Spain}

\author{A. Pastore}
\email{alessandro.pastore@york.ac.uk}
\affiliation{Department of Physics, University of York, Heslington, York, Y010 5DD, United Kingdom}


\begin{abstract}
We present the development of the extended Skyrme N2LO pseudo-potential in the case of spherical even-even nuclei calculations. The energy density functional is first presented. Then we derive the mean-field equations and discuss the numerical method used to solve the resulting fourth-order differential equation together with the behaviour of the solutions at the origin. Finally, a fitting procedure for such a N2LO interaction is discussed and we provide a first parametrization. Typical ground-state observables are calculated and compared against experimental data. 
\end{abstract}


\pacs{
    21.30.Fe 	
    21.60.Jz 	
}
 
\date{\today}


\maketitle


\section{Introduction}
\label{sect:intro}

The Nuclear Energy Density Functional (NEDF) theory allows us to describe properties of nuclei from light to heavy nuclei and from drip-line to drip-line~\cite{ben03}.
Several functionals have been developed in the recent years, but  the most widely used~\cite{per04,rai11} are those derived from the non-relativistic zero-range Skyrme interaction~\cite{sky59}. Since its first applications to atomic nuclei~\cite{vau72}, this interaction has proven to be very well suited to describe nuclear observables at very reduced computational cost~\cite{gor09}. 

A  crucial aspect in building a functional is to determine the values of its coupling constants. Despite its apparent simplicity, this is a very delicate aspect: a badly determined coupling constant can give rise to unphysical instabilities~\cite{les06,sch10,frac12,hel13,Pas15T,report,dep16} and thus to unphysical results. 
A possibility for avoiding them is to find an adequate set of observables so that all coupling constants are properly constrained during the optimization procedure~\cite{dob14,nik16}. In Ref.~\cite{pas13}, we have presented an alternative solution to avoid unphysical instabilities based on the linear response (LR) formalism in infinite nuclear medium. This solution is particularly simple and very efficient especially for some particular terms of the functional that are odd under time reversal symmetry and give very little contribution to masses of odd-systems~\cite{sch10}.  However, avoiding unphysical instabilities is not the only requirement to have an effectient functional : one also has to check how it performs to describe nuclear observables. On this point, the  UNEDF collaboration~\cite{ber07} has recently studied much in detail the properties of Skyrme functionals against a large set of nuclear observables~\cite{kor10,kor12,kor13}. The main conclusion in their last article~\cite{kor13} is that  the standard Skyrme functional~\cite{per04} has reached its limits. If we want to improve the description of experimental data (as masses, radii, fission barriers,...) we need to follow two paths: explore different functional forms or develop functionals at multi-reference level~\cite{dug15}.

Following the idea of Carlsson and collaborators~\cite{car08,rai11}, we have decided to explore the first path and to study the impact of additional gradient terms into the Skyrme pseudo-potential~\cite{dav13}. The gradient terms have been introduced in a systematic way by considering all possible combinations allowed by the symmetries of the problem up to 6th power. The resulting pseudo-potential has been called N$\ell$LO which by definition incorporates gradients up to order $2\ell$. 
Within this language, the standard Skyrme interaction~\cite{cha97} is named N1LO. In Ref.~\cite{Dav16AN}, we have shown 
the explicit connection between the Taylor momentum expansion of  $any$ finite range interaction and the actual form of the N$\ell$LO pseudo-potential~\cite{rai11}.
In that article, we have also proven that such an expansion works fairly well in infinite nuclear medium and that the main properties of the Equation of State (EoS) of a finite-range interaction can be fairly reproduced by truncating the momentum expansion to fourth order (N2LO). The result is coherent with previous findings based on Density Matrix Expansion (DME)~\cite{car10}: the role of fourth order terms is important and it leads to a remarkable improvement of the DME results when compared to finite-range interactions. Higher order terms can thus be neglected as a first step since their contribution becomes systematically less important.

At present, the only existing parametrizations of the extended Skyrme N2LO/N3LO pseudo-potentials have been obtained by considering only properties of infinite nuclear medium~\cite{Dav15,Dav15AA}, that is without taking into account properties of finite nuclei. In order to remedy this aspect, we present here a new  Skyrme Hartree-Fock-Bogoliubov (HFB) code that incorporates higher order derivatives terms appearing in N2LO. It is worth remainding at this point that an alternative  code named HOSPHE~\cite{hosphe} has already been published. This code, based on Harmonic-Oscillator (HO) basis also considers the most general functional form of the N3LO functional~\cite{car08} using spherical basis representation. However, following our previous findings of Ref.~\cite{dav13}, we have decided to express the N$\ell$LO pseudo-potential in Cartesian coordinates and to develop for this specific case a numerical code to work in coordinate space: the r-space representation is in fact more convenient to be used in a fitting procedure since we do not need to use a very large number of basis states to achieve convergence. See Ref.~\cite{sch15} for more details.

The article is organized as follows: in Sec.~\ref{sec:n2lo} we present the general functional formalism for the N2LO pseudo-potential and in Sec.~\ref{sec:n2lo:spheric} we specialize the formalism for the spherically symmetric case. In Sec.~\ref{sec:hfb} we present in detail the generalization of the Hartree-Fock-Bogoliubov equations to include the N2LO pseudo-potential. In Sec.~\ref{sec:fit} we present the fitting protocol to determine the parameters of the new N2LO functionals. Finally we give our conclusions in Sec.~\ref{sec:conclusions}.

\section{N2LO Skyrme functional}\label{sec:n2lo}

The N2LO Skyrme pseudo-potential as described in Refs.~\cite{car08,rai11} is a generalization of the standard Skyrme interaction, corresponding to the expansion of the momentum space matrix elements of a generic interaction in powers of the relative momenta $\mathbf{k}, \mathbf{k}'$ up to the fourth order. Following~\cite{dav14c}, the form considered in this article respects both Galilean and local gauge invariance~\cite{dob95}. It is written as the sum of three terms
\begin{eqnarray}
V_{\text{N2LO}} =V_{\rm N2LO}^{C}+V_{\rm N1LO}^{LS}+V_{\rm N1LO}^{DD}\;.
\end{eqnarray}
The central term reads
\begin{eqnarray} \label{eq:N2LO:c}
V_{\rm N2LO}^{C} &=& t_0 (1+x_0 P_{\sigma}) \nn \\
            && + \frac{1}{2} t_1 (1+x_1 P_{\sigma}) ({\mathbf{k}}^2 + {\mathbf k'}^2)  \nn \\
            &&  + t_2 (1+x_2 P_{\sigma}) ({\mathbf k} \cdot {\mathbf k'})  \nonumber\\
            & & + \frac{1}{4} t_1^{(4)} (1+x_1^{(4)} P_{\sigma}) \left[({\mathbf k}^2 + {\mathbf k'}^2)^2 + 4 ({\mathbf k'} \cdot {\mathbf k})^2\right] \nonumber\\
            & &+ t_2^{(4)} (1+x_2^{(4)} P_{\sigma}) ({\mathbf k'} \cdot {\mathbf k}) ({\mathbf k}^2 + {\mathbf k'}^2) .
\end{eqnarray}
In these expressions, a Dirac function $\delta({\mathbf r}_1-{\mathbf r}_2)$
is to be understood, but has been omitted for the sake of clarity. See Ref.~\cite{ben03} for details on the adopted notations. The spin-orbit term $V_{\rm N1LO}^{LS}$ is not affected by the inclusion of higher order gradient terms: in Ref.~\cite{Dav16AN}, we have shown that other possible spin-orbit terms are suppressed once the local gauge invariance~\cite{rai11,rai11b} is imposed.  In Ref.~\cite{Dav16AN}, we have discussed in details the problem of local gauge invariance for spin-orbit term and in particular the possible violation of such a symmetry for finite-range spin-orbit terms. The density-dependent term $V_{\rm N1LO}^{DD}$ has also exactly the same structure as in the standard Skyrme interaction~\cite{cha97}, since its nature is to mimic the effect of a three-body term~\cite{vau72,sad13}. Tensor terms should be also included into Eq.~(\ref{eq:N2LO:c}). In Ref.~\cite{Dav15}, we have discussed them based on the partial-wave decomposition of the total EOS. In finite nuclei it is actually very difficult to constrain them in NEDF~\cite{sag14} because of their strong competition with the spin-orbit term in modifying the underlying single-particle structure~\cite{les07}. For this preliminary exploration, we have thus decided to neglect them. Finally, it is worth mentioning that in the present article we will always use the complete interaction in the sense that we will not discard the so-called $J^2$ tensor terms~\cite{les07} as often done in the literature. For the Coulomb interaction between protons, we adopt the same procedure as described in Ref.~\cite{cha97} \emph{i.e.} using the standard Slater approximation for the exchange term~\cite{ska01}.

Starting from Eq.~(\ref{eq:N2LO:c}), it is possible to derive the explicit form of the Skyrme functional in Cartesian coordinates. We write it as
\begin{eqnarray}\label{eq:func:gen}
\mathcal{E}=\sum_{t}\mathcal{E}^{(1),\text{even}}_t +\mathcal{E}^{(1),\text{odd}}_t +\mathcal{E}^{(2),\text{even}}_t +\mathcal{E}^{(2),\text{odd}}_t \;,
\end{eqnarray}
where $t=0,1$ is the isospin index and even/odd refers to the behaviour of the terms of the functional under time-reversal symmetry~\cite{per04}. In the above equation, we have explicitly separated the contributions originated from the N$\ell$LO terms $\mathcal{E}^{(\ell=1,2)}$ . The standard terms $\mathcal{E}^{(1)}_t$ read~\cite{les07}
\bqr
\mathcal{E}^{(1),\text{even}}_t & = &
        C_t^\rho [\rho_0 ] \, \rho_t^2
      + C_t^{\Delta \rho}  \, \rho_t \, \Delta \rho_t
      + C_t^\tau          \, \rho_t \, \tau_t
      - C^{T}_t \sum_{\mu, \nu = x}^{z} J_{t,\mu \nu} J_{t,\mu \nu}  +C_t^{\nabla J} \; \rho_t \, \nabla \cdot {\bf J}_t  \,,      \\
\label{eq:centEDFo}
\mathcal{E}^{(1),\text{odd}}_t & = &
        C_t^s [\rho_0 ] \, {\bf s}_t^2
      - C_t^\tau        \, {\bf j}^2_t
      + C^{\Delta s}_t  \, {\bf s}_t \cdot \Delta {\bf s}_t
      + C^{T}_t         \, {\bf s}_t \cdot {\bf T}_t     +            C^{\nabla J}_t \; {\bf s}_t \cdot \nabla \times {\bf j}_t  \,,  
\eqr
while the new terms can be written as
\begin{eqnarray}
\mathcal{E}_t^{\text{(2),even}}
& = & C^{( \Delta \rho)^2}_t \left( \Delta \rho_t \right)^2
    + C^{ M \rho}_t \bbM_t^{M \rho,\text{even}} 
    + C^{ M s}_t \bbM_t^{Ms,\text{even}}              \, ,   \\
\label{eq:ef:DKo}
\mathcal{E}_t^{\text{(2),odd}}
& = & C^{(\Delta s)^2}_t \left( \Delta \bs_t \right)^2
    + C^{ M \rho}_t \bbM_t^{M \rho,\text{odd}} 
    + C^{ M s}_t \bbM_t^{Ms,\text{odd}}               \, ,
\end{eqnarray}
where
\bqr
\bbM^{M \rho,\text{even}} 
& = &  \left\{ \, \rho \, Q  \, + \, \tau^2 \, \right\}  + 2 \,\left[ \mbox{Re}(\tau_{\mu \nu}) \mbox{Re}(\tau_{\mu \nu}) 
   \, - \, \mbox{Re}(\tau_{\mu \nu}) \nabla_{\mu} \nabla_{\nu} \rho \;  \right]                  \, ,  \\
\label{taumunuH}
\bbM^{Ms,\text{even}} 
& = &  \left\{ \, \left( \nabla_{\mu} J_{\mu \nu} \right)^2
   \, + \, 4 J_{\mu \nu} V_{\mu \nu} - \mbox{Im}(K_{\mu \nu \kappa}) \mbox{Im}(K_{\mu \nu \kappa}) \, \right\}                           \, ,     \\
\bbM^{M \rho,\text{odd}} 
& = &  \left\{ \, \left( \vnabla \cdot \bj \right)^2
   \, + \, 4 \, \bj \cdot \vPi \, \right\}                                 \, ,     \\
\bbM^{Ms,\text{odd}} 
& = &  \left\{ \, \bs \cdot \bS  \, + \, \bT^2  \, \right\} +  2 \, \left[ \mbox{Re}(K_{\mu \nu \kappa}) \mbox{Re}(K_{\mu \nu \kappa}) - \mbox{Im}(\tau_{\mu \nu})\mbox{Im}(\tau_{\mu \nu})
   \, - \, \mbox{Re}(K_{\mu \nu \kappa}) \nabla_{\mu} \nabla_{\nu} s_{\kappa} \right].
\label{KmunuH}
\eqr
These terms contain six new densities:  $\tau_{\mu \nu}$, $V_{\mu \nu}$, $\mathbf{\Pi}, K_{\mu\nu\kappa}, Q$ and $\mathbf{S}$. Their explicit definition is given in Appendix~\ref{app:dens}.


\section{N2LO functional in spherical symmetry}\label{sec:n2lo:spheric}

In the present section, we limit ourselves to the case of spherical symmetry. In this case, the single-particle wave function can be written as follows
\bqr
\psi_{n \ell j m q} (\br)
            =  \frac{1}{r} R_{n\ell j q} (r) \; \Omega_{\ell j m} ({\hat r})  \, ,
\label{fosphe}
\eqr
where $n$ is the principal quantum number,  $\Omega_{\ell j m} ({\hat r})$ is a solid spherical harmonic~\cite{var88} and $\ell j m$ refer respectively to the orbital angular momentum, the total angular momentum and its relative projection along the $z$-axis. Here $q\equiv n,p$ stands for proton (p) or neutron (n). In our formalism the two nuclear species are not mixed explicitly~\cite{per04,sat13}. By considering only even-even systems, we can further simplify the expressions given in Eq.~(\ref{eq:func:gen})
\bqr
\label{eq:EDF_N1LO_C_sphere}
\mathcal{E}^{(1)}
   = &&   C^{\rho}_0 \, \rho_0^2
 \, +  \, C^{\rho}_1 \, \rho_1^2
 \, + \,  C^{\Delta \rho}_0  \, \rho_0 \Delta \rho_0
 \, + \,  C^{\Delta \rho}_1 \, \rho_1 \Delta \rho_1                                  \\
  & + &   C^{\tau}_0  \, \rho_0 \tau_0
 \, + \,  C^{\tau}_1 \, \rho_1 \tau_1
 \, - \,   \tfrac{1}{2}\,  C^T_0  \, J_0^2
 \, - \,  \tfrac{1}{2}\, C^T_1 \, J_1^2                                                              \nn \\
  & + &  C^{\nabla J}_0\, \rho_0 \, \vnabla \cdot \bJ_0
 \, + \,  C^{\nabla J}_1 \, \rho_1 \, \vnabla \cdot \bJ_1             \;;      \nn
\eqr
\begin{eqnarray}
\label{eq:EDF_N2LO_even_cc_C}
\qquad \qquad \qquad \qquad  \mathcal{E}^{(2)}
  =&&  C^{(4) \Delta \rho}_0 \, \left( \Delta \rho_0 \right)^2   \, + \, \, C^{(4) \Delta \rho}_1 \, \left( \Delta \rho_1 \right)^2                        \nn  \\
  & +&   C^{(4) M \rho}_0
  \, \Big\{ \, \left[ \, \rho_0 \, Q_0  \, + \, \tau_0^2 \, \right] \Big\} \nn \\
  & +&  C^{(4) M \rho}_0 \, \Big\{\, \left[ \mbox{Re}(\tau_{0, \mu \nu}) \mbox{Re}(\tau_{0, \mu \nu}) 
 \, - \mbox{Re}(\tau_{0, \mu \nu}) \nabla_{\mu} \nabla_{\nu} \rho_0 \right] \, \Big\}              \nn \\
  & + &  \, C^{(4) M \rho}_1 \, \Big\{ \, \left[ \, \rho_1 \, Q_1 \, + \, \tau_1^2 \, \right] \Big\} \nn \\
 & +&  C^{(4) M \rho}_1 \, \Big\{\left[ \mbox{Re}(\tau_{1, \mu \nu}) \mbox{Re}(\tau_{1, \mu \nu}) 
 \, - \,  \mbox{Re}(\tau_{1, \mu \nu}) \nabla_{\mu} \nabla_{\nu} \rho_1 \right] \, \Big\}             \nn \\
  & -  &  C^{(4) M s}_0
                      \, \left[ \, \left( \nabla_{\mu} J_{0, \mu \nu} \right)^2
 \, + \, 4 J_{0, \mu \nu} V_{0, \mu \nu} - \mbox{Im}(K_{0,\mu \nu \kappa}) \mbox{Im}(K_{0,\mu \nu \kappa}) \, \right]       \nn \\
& - &   C^{(4) M s}_1 \, \left[ \, \left( \nabla_{\mu} J_{1, \mu \nu} \right)^2
 \, + \, 4 J_{1, \mu \nu} V_{1, \mu \nu} - \mbox{Im}(K_{1,\mu \nu \kappa}) \mbox{Im}(K_{1,\mu \nu \kappa}) \, \right] \,.
\label{N2LOfonct}
\end{eqnarray}

\subsection{Local densities}

Let us introduce the short-hand notation $\alpha=\{n\ell jq\}$ and $C_\alpha = j(j+1) - \ell(\ell+1) - \frac{3}{4}$.
The explicit expressions of the densities in spherical symmetry (we limit ourselves to systems that are even under time-reversal) up to second order take the form~\cite{ben05}
\begin{eqnarray}
\rho_{0} (r)               & = & \ \sum_{\alpha} \  \frac{(2 j + 1 )}{4 \pi} \ \frac{R_{\alpha}^2(r)}{r^2} \,, \\
\tau_{0} (r)               & =&  \ \sum_{\alpha} \  \frac{(2 j + 1 )}{4 \pi r^2} \ \left[ \left(R_\alpha^\prime(r) - \frac{R_\alpha(r)}{r} \right)^2 + \frac{\ell (\ell+1)}{r^2} R_{\alpha}^2(r) \right]  \,, \\
 \label{Jraddens} J_0(r) &=&  \sum_{\alpha} \frac{(2 j + 1 )}{4 \pi}    \mbox{C}_\alpha \frac{R_{\alpha} (r)}{r^3} \,.
\end{eqnarray}
$\tau_{0} (r) $ can be conveniently decomposed in a radial and centrifugal part as $\tau_0=\tau_{R,0}+\tau_{C,0}$ where
\begin{eqnarray}
\tau_{R,0}(r)&=&\  \sum_\alpha \frac{(2 j + 1 )}{4 \pi r^2} \left [ R_{\alpha}^\prime (r)-\frac{R_{\alpha}(r)}{r}\right] ^2 \,,\\
\tau_{C,0} (r) &= & \sum_{\alpha} \frac{(2 j + 1 )}{4 \pi}   \frac{\ell ( \ell + 1)}{r^2} \frac{R^2_{\alpha} (r)}{r^2}     \,.
\end{eqnarray}
Eq.~(\ref{Jraddens}) corresponds to the radial part of the $J_{\mu\nu,0} (r)$ spin-orbit vector density defined as
\bqr
\label{eq:Jmunu}
J_{\mu \nu,0} & = &  \frac{1}{2} \, \epsilon_{\mu \nu \kappa} \, J_{\kappa}    
           \, = \, \frac{1}{2} \, \epsilon_{\mu \nu \kappa} \, \frac{X_\kappa}{r} \, J_0(r)   \, ,
\eqr
where $X_{\mu}$ represents the Cartesian coordinates. If we now come to fourth order, the explicit expressions of the new densities in spherical symmetry take the form
\begin{eqnarray}
\tau_{\mu\nu, 0}   (r)  &=&  \frac{1}{2} \ \tau_{C,0}(r) \; \delta_{\mu\nu} \, + \, \frac{X_\mu X_\nu}{r^2} \left[\tau_{R,0}(r) - \frac{1}{2}  \ \tau_{C,0}(r)\right]  \;,  \\
  V_{0}(r)&=& \sum_{\alpha} \frac{(2 j + 1 )}{4 \pi r^2} \ \mbox{C}_\alpha  \left[ \frac{ R_{\alpha}^2}{r^3} \ \left[  \ell(\ell+1) +2  \right] +\frac{R^{\prime2}_{\alpha}(r)}{r} -4\frac{R^{\prime}_{\alpha}(r)R_{\alpha}(r) }{r^2}\right] \; ,\\
Q_{0}(r) &=&  \sum_{\alpha} \frac{(2 j + 1 )}{4 \pi r^2} \left[R^{\prime\prime}_{\alpha}(r)- \ell(\ell+1)\frac{R_{\alpha}(r)}{r^2}\right]^2 \;,\\
 K_{\mu \nu \kappa, 0}(\br) &=& i \ \mbox{K1}_0 (r) \epsilon_{\mu \nu \kappa} + i \ \mbox{K2}_0 \left[ \epsilon_{\mu \kappa M} \frac{X_M X_\nu}{r^2} + \epsilon_{\mu \nu M} \frac{X_M X_\kappa}{r^2} + \epsilon_{\kappa \nu M} \frac{X_M X_\mu}{r^2} \right] .  
\end{eqnarray}
We have defined $K1_{0}$ and  $K2_{0}$ as
\begin{eqnarray}
 \label{K_comp}
  \mbox{K1}_0 (r) &=& \sum_\alpha \frac{(2 j + 1 )}{16 \pi r^3} \mbox{C}_\alpha R_\alpha^{\prime}(r) R_\alpha(r) \;,\\
  \mbox{K2}_0 (r) &=& \sum_\alpha \frac{(2 j + 1 )}{16 \pi r^3} \mbox{C}_\alpha \left[ \frac{2}{r} R_\alpha(r)^2 - R_\alpha^{\prime}(r) R_\alpha(r) \right].
  \end{eqnarray}
$\tau_{\mu\nu, 0}$(r) is the kinetic density tensor. The usual N1LO $\tau_0$(r) density is given by its trace
\begin{equation}
\sum_{\mu} \tau_{\mu\mu,0}(r) = \tau_0.
\end{equation}
The even part of the N2LO functional only receives a non-vanishing contribution from the real part of this density (Eq.~\ref{N2LOfonct}). Given that the imaginary part is zero under spherical symmetry, we will write $\tau_{\mu\nu, 0}$(r) instead of Re($\tau_{\mu\nu, 0}$(r)) in the following.
Similarly to  $J_{0}(r)$, $V_{0}(r)$ is the radial part of the vector density $V_{\mu \nu, 0}$
\begin{equation}
V_{\mu \nu, 0}= \frac{1}{2} \ \epsilon_{ \mu \nu \kappa} \  \frac{X_\kappa}{r} \  V_0 (r),
\label{Vmunusph}
\end{equation}
and it can be decomposed in a radial and centrifugal part as $V_0 =V_{R,0}+V_{C,0}$ where
\begin{eqnarray}
 V_{R,0}(r) &=& \sum_{\alpha} \frac{(2 j + 1 )}{4 \pi r^3} \mbox{C}_\alpha  \left[ R_\alpha^{\prime^2}(r) - \frac{4}{r} R_\alpha^{\prime}(r) R_\alpha(r) + \frac{2}{r^2} R_\alpha^2(r) \right] . \\
 V_{C, 0}(r) &=& \sum_{\alpha} \frac{(2 j + 1 )}{4 \pi r^3} \mbox{C}_\alpha  \left[  \frac{\ell(\ell+1)}{r^2} R_\alpha^2(r)  \right] .
\end{eqnarray}
Since the $K_{\mu \nu \kappa, 0}(\br)$ density is imaginary in spherical symmetry, the N2LO functional (Eq.~\ref{N2LOfonct}) only receives a contribution of this density multiplied by itself.
As for the $\tau_{\mu\nu, 0}$(r), we will use $K_{\mu \nu \kappa, 0}(\br)$ without mentioning anymore that it actually stands for the imaginary part of this density. \\

\begin{figure}[!h]
\begin{center}
\includegraphics[width=0.5\textwidth]{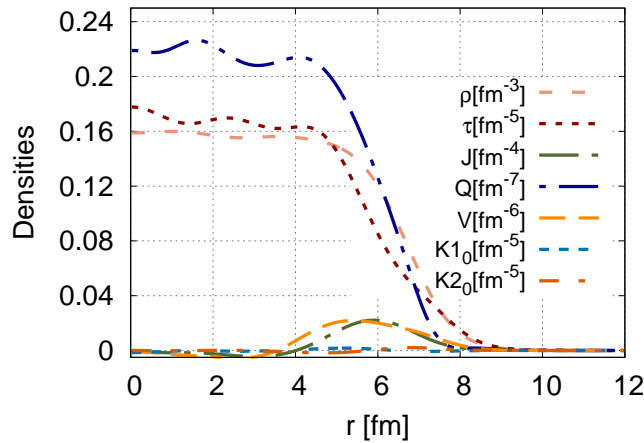}
\end{center}
\caption{(Colors online) Isoscalar densities in $^{208}$Pb calculated using single particle wave functions obtained by a SLy5 mean-field solution. See text for details. }
\label{WS:density}
\end{figure}

Some additional expressions which represent the new contributions to the functional are also written below for completeness.
\begin{eqnarray}
\tau_{\mu \nu, 0}(r) \tau_{\mu \nu, 0} (r)&=& \tau_{R,0}^2 (r)- \frac{1}{2} \tau_{C,0}^2(r)
\end{eqnarray}
\begin{equation}
\tau_{\mu \nu,0} \nabla_\mu \nabla_\nu \rho =\rho^{(2)} \tau_R + \frac{\rho^{(1)}}{r} \tau_C
\end{equation}
\begin{eqnarray}
J_{\mu \nu,0} V_{\mu \nu,0} &=& \frac{1}{2 } J_0 (r) V_0 (r)
\end{eqnarray}
 \begin{equation}\label{k2fonct}
 K_{\mu \nu \kappa,0} K_{\mu \nu \kappa,0} =  6  \mbox{K1}_{0}(r)^2 +  6  \mbox{K2}_{0}(r)^2 - 4 \mbox{K1}_{0}(r) \mbox{K2}_{0}(r).
 \end{equation}
In order to have a qualitative and quantitative idea of all these densities, we represent in Fig.~\ref{WS:density}, the isoscalar  densities in $^{208}$Pb. These densities have been determined using a single particle basis obtained from a fully-converged Hartree-Fock (HF) solution based on the SLy5 functional~\cite{cha97}.
We observe that all the densities used here are well-behaved at the origin of the coordinate system.

\section{Hartree-Fock-Bogoliubov equations in spherical symmetry}\label{sec:hfb}

In this section we describe the method used to solve the complete Hartree-Fock-Bogoliubov (HFB) equations and the numerical tests we have performed. 

\subsection{Hartree-Fock}

We start considering closed-shell nuclei for which the HFB equations can be safely reduced to the standard Hartree-Fock (HF) equations. They read~\cite{vau72,rin80}
\begin{eqnarray}\label{eq:hf}
h_q(r) R_{nljq}(r)=\varepsilon^q_{nlj} R_{nljq}(r)\;,
\end{eqnarray}
where $R_{nljq}(r)$ is the radial part of the single-particle wave-function given in Eq.~(\ref{fosphe}). The corresponding Hamiltonian is derived as a functional derivative as
\begin{eqnarray}\label{sp:eq:4th}
h_q(r)  &=&A^q_4 \frac{d^4}{dr^4} +  A^q_3 \frac{d^3}{dr^3} + A^q_{2 R}  \frac{d^2}{dr^2} + A^q_{1 R}  \frac{d}{dr} + A^q_{0 R} \nonumber\\
  &+& \frac{ \ell (\ell+1)}{r^2} \left[ A^q_{2 C}  \frac{d^2}{dr^2} + A^q_{1 C} \frac{d}{dr}  + \frac{ \ell (\ell+1)}{r^2} A^q_{0 CC} + A^q_{0 C}   \right] \nn \\
   &+& W^q_{2 R}  \frac{d^2}{dr^2} +  W^q_{1 R}  \frac{d}{dr} +    W^q_{0 R} + \frac{ \ell (\ell+1)}{r^2}  W^q_{0 C}    \;.
\label{eqndiff4}
\end{eqnarray}
We observe that the inclusion of 4th order term in the interaction translates into a fourth order differential equation. Although this is quite unusual in nuclear physics, a 4th order differential equation is routinely solved in other physical systems, as for example to describe the behaviour of a bending solid beam~\cite{ban01}.

\begin{figure}[!h]
\begin{center}
\includegraphics[width=0.45\textwidth,angle=0]{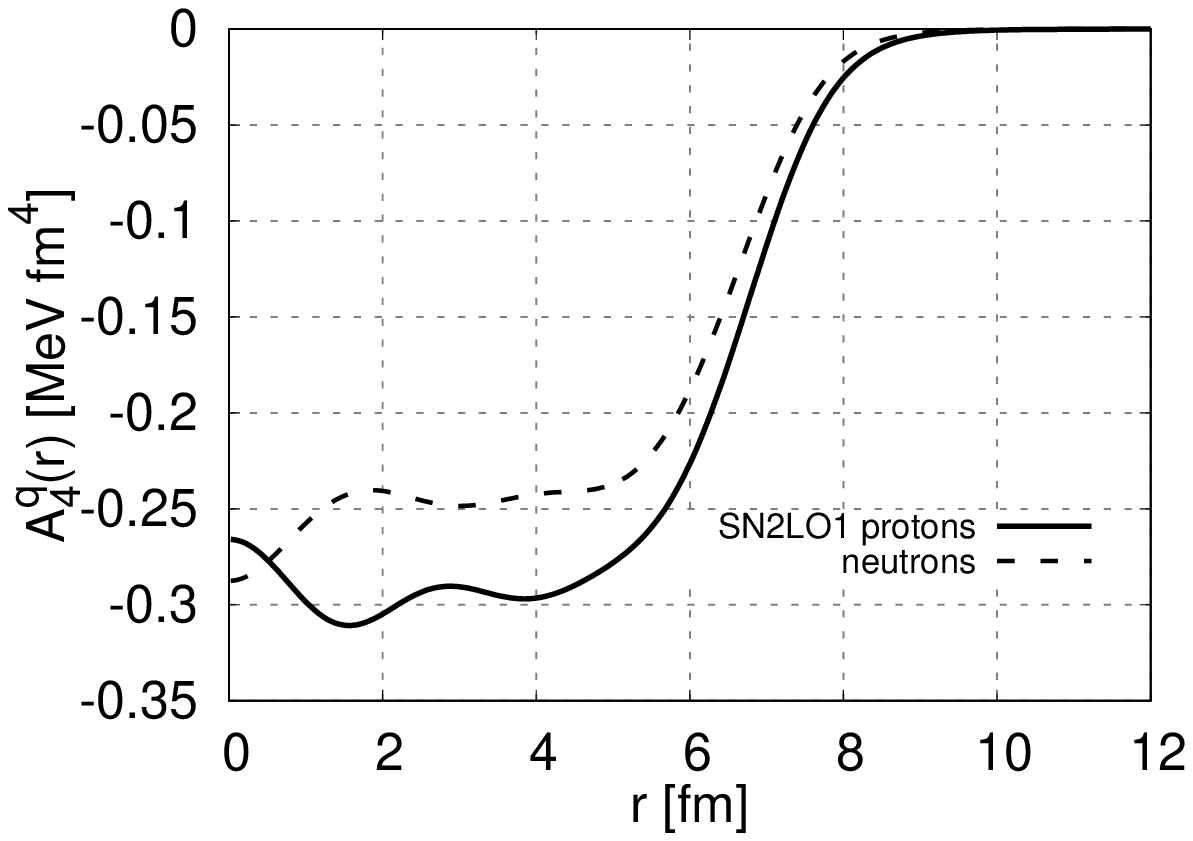}
\includegraphics[width=0.45\textwidth,angle=0]{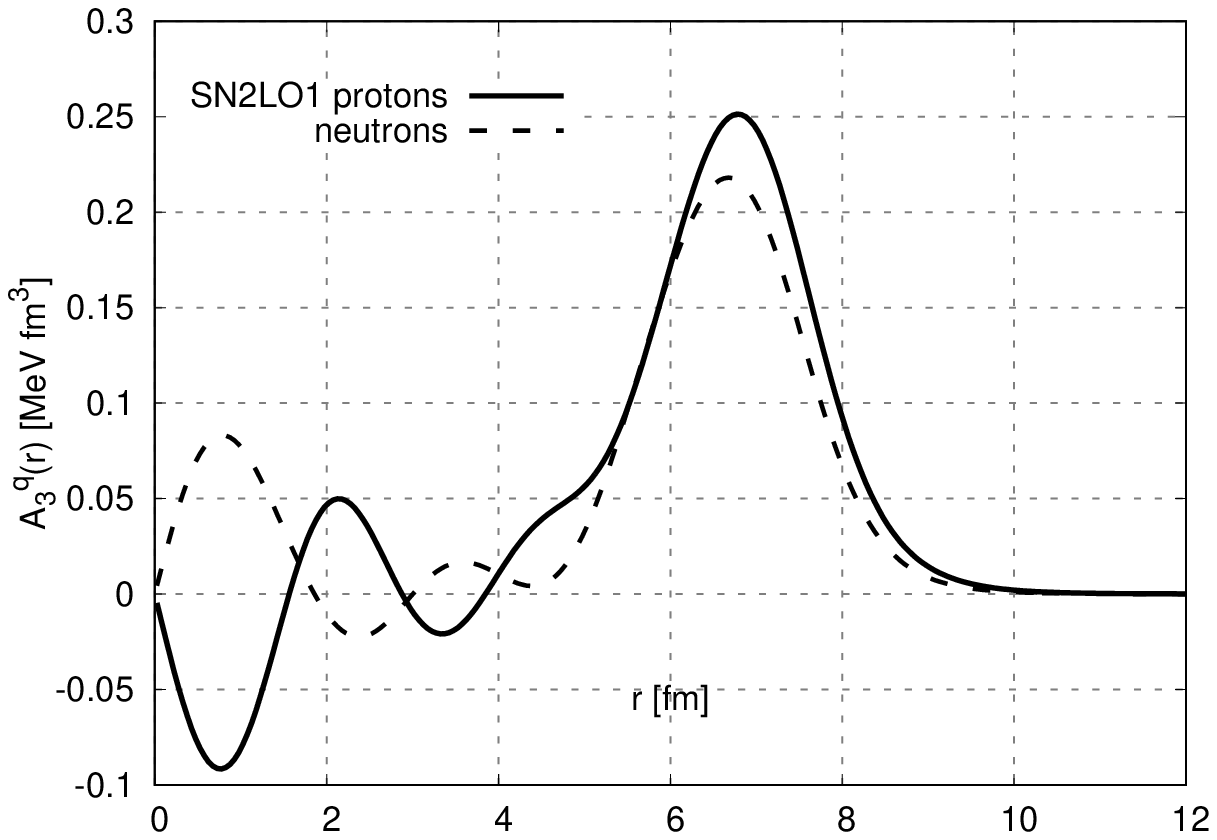}\\
\includegraphics[width=0.45\textwidth,angle=0]{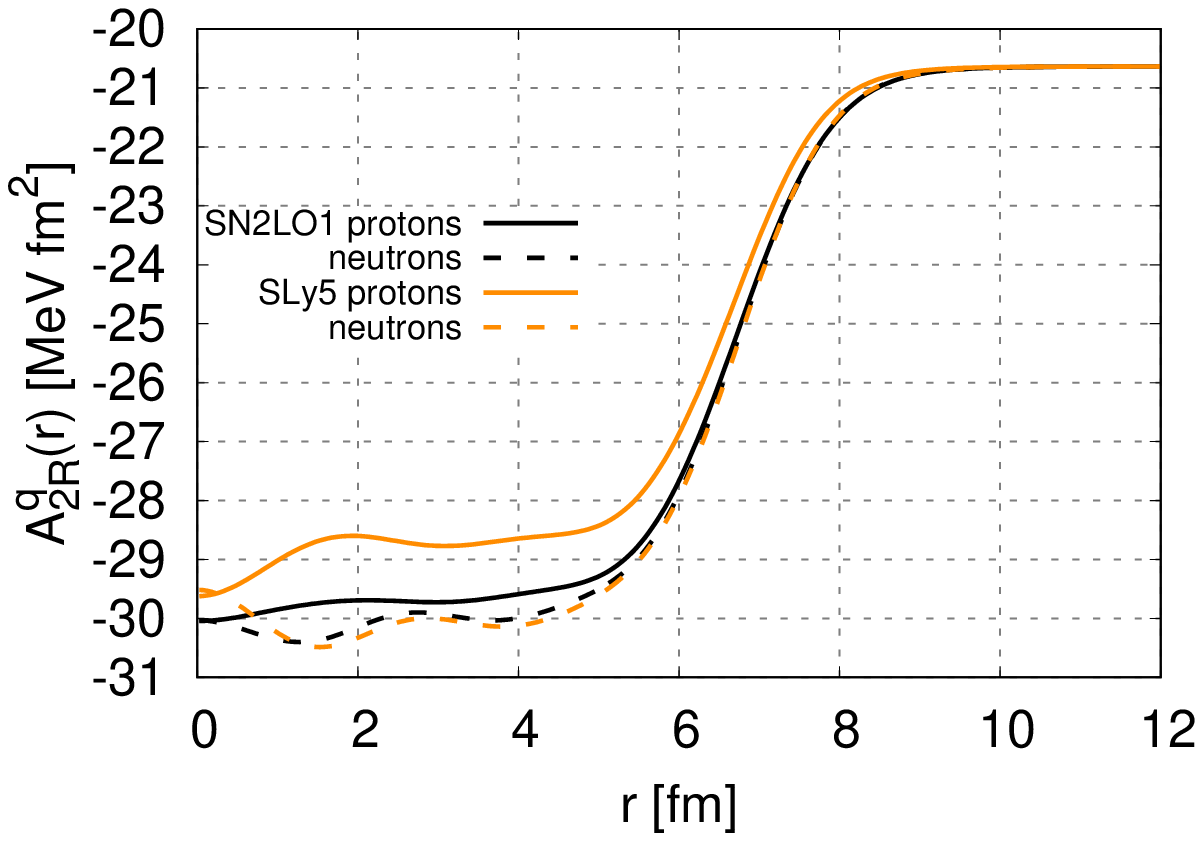}
\includegraphics[width=0.45\textwidth,angle=0]{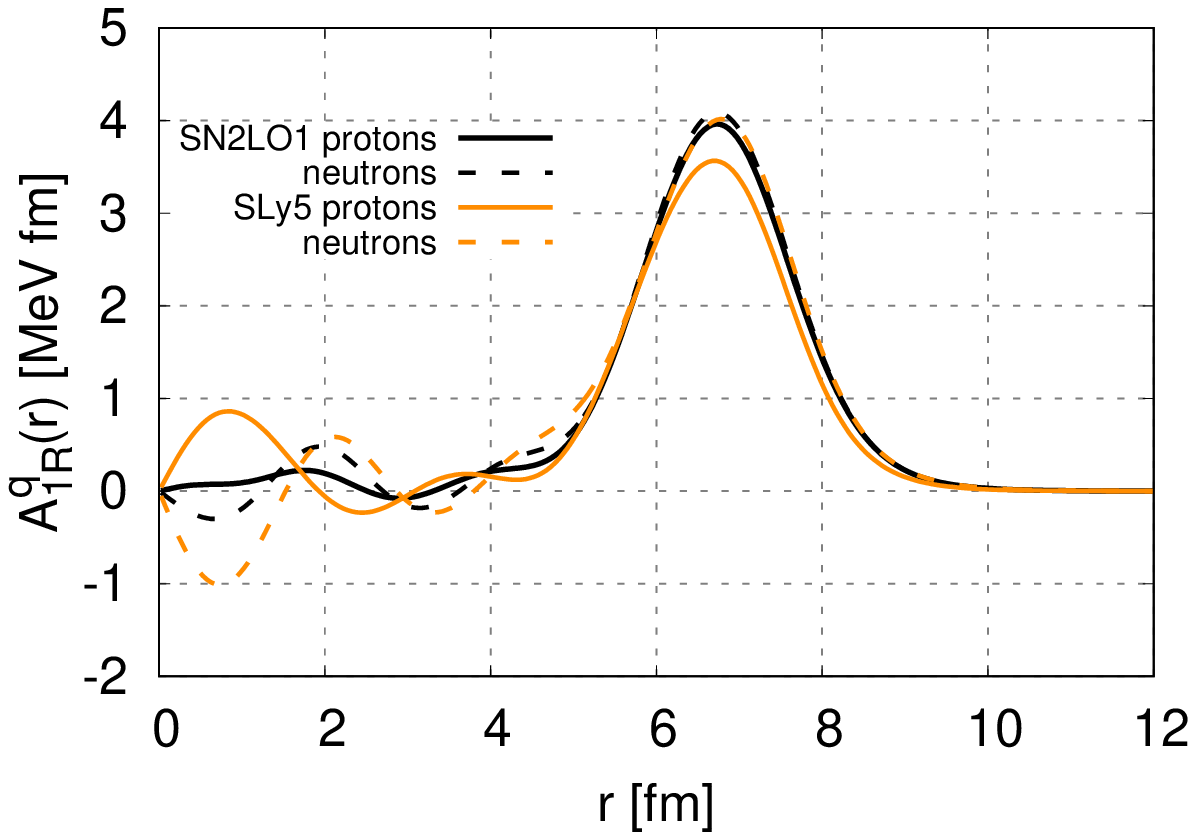}
\includegraphics[width=0.45\textwidth,angle=0]{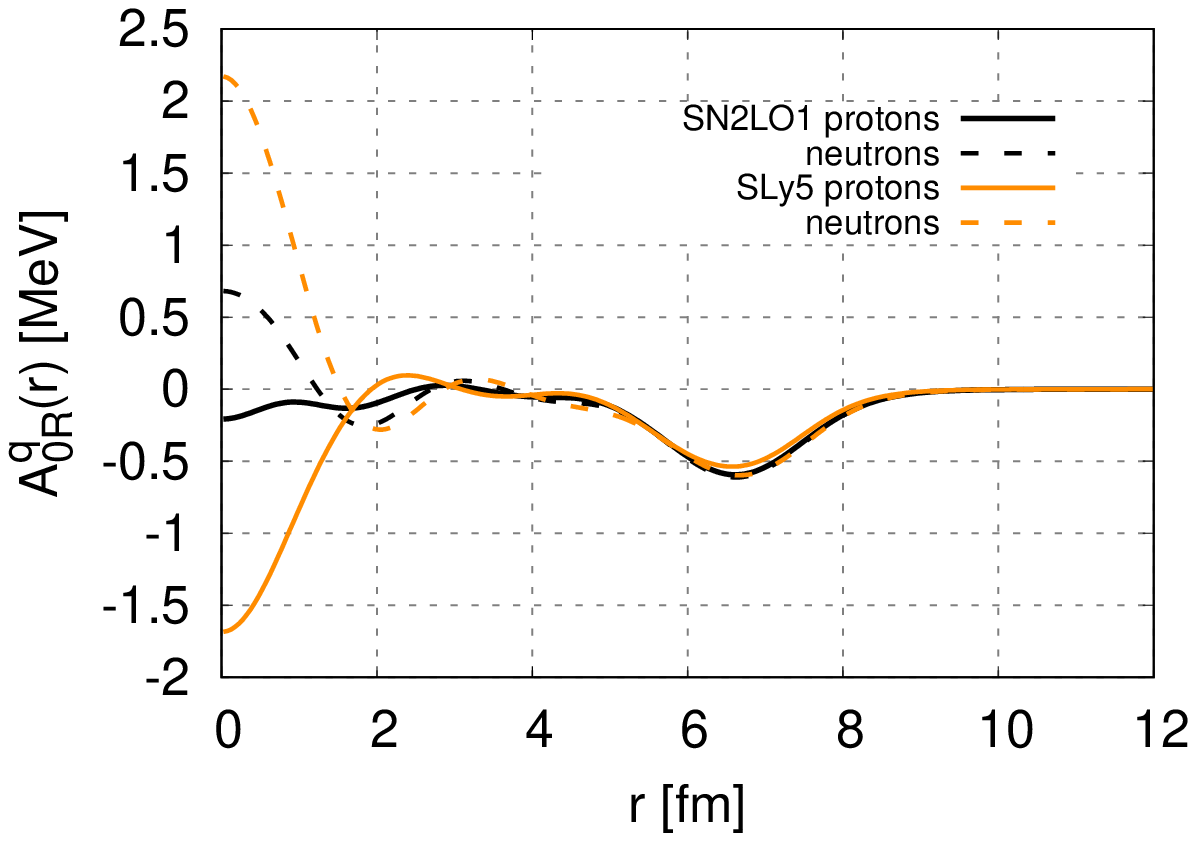}
\includegraphics[width=0.45\textwidth,angle=0]{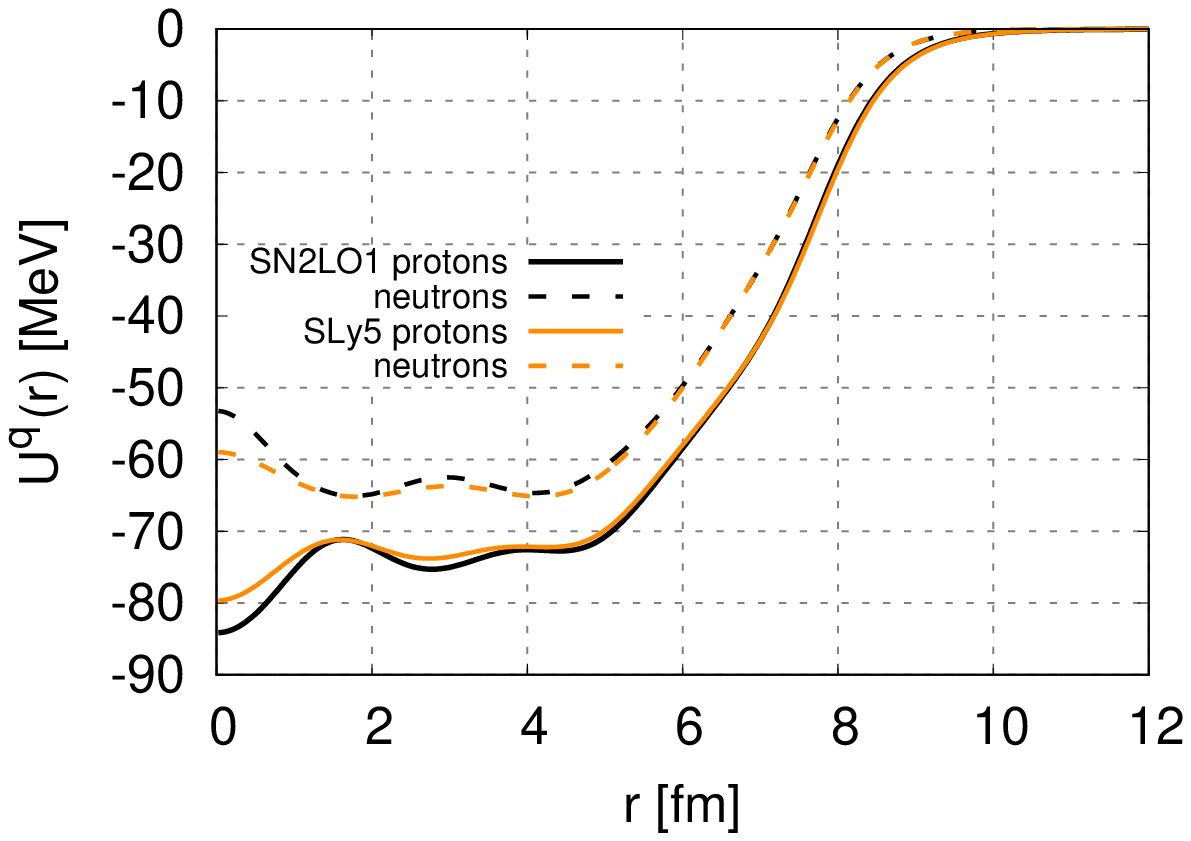}
\end{center}
\caption{(Colors online) Radial dependence of the coefficients defined in Eq.~(\ref{sp:eq:4th}) for $^{208}$Pb obtained using the SN2LO1 and SLy5 interactions. See text for details.}
\label{208pb:field}
\end{figure}

\begin{figure}[!h]
\begin{center}
\includegraphics[width=0.45\textwidth,angle=0]{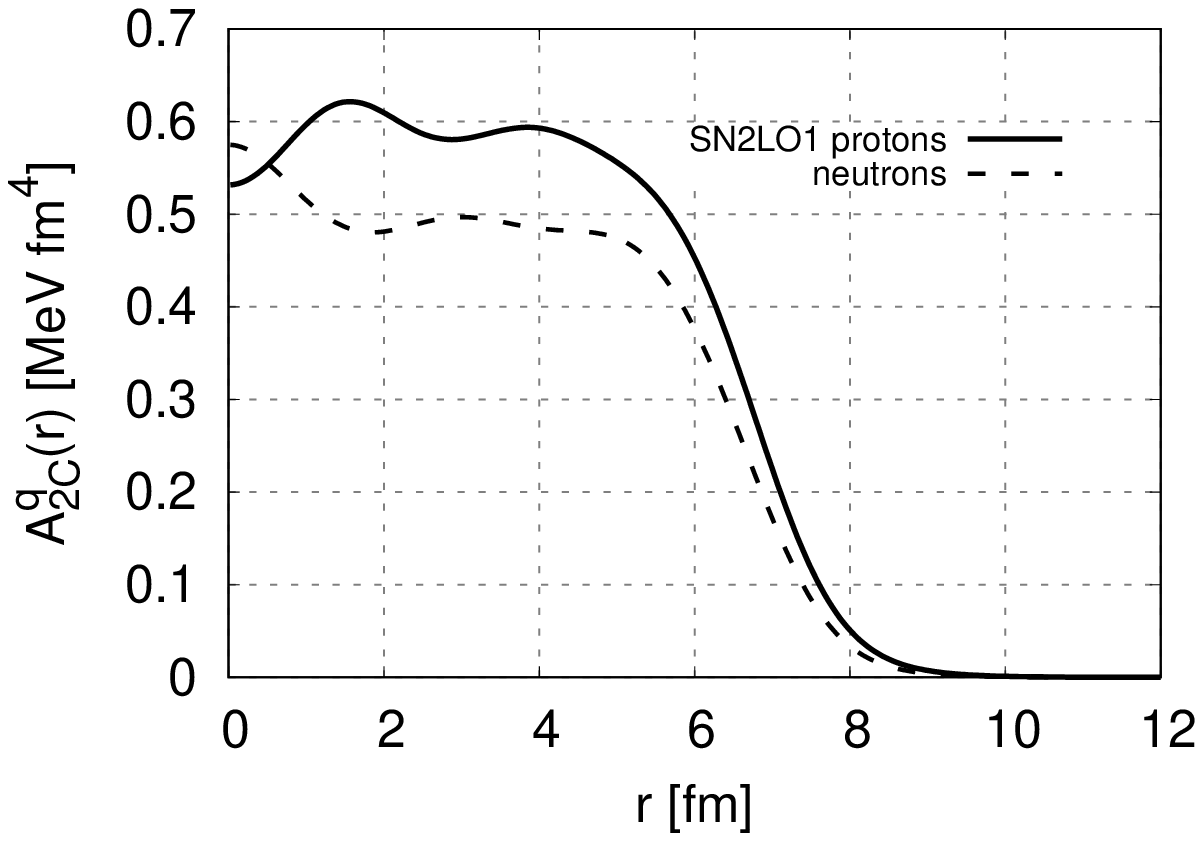}
\includegraphics[width=0.45\textwidth,angle=0]{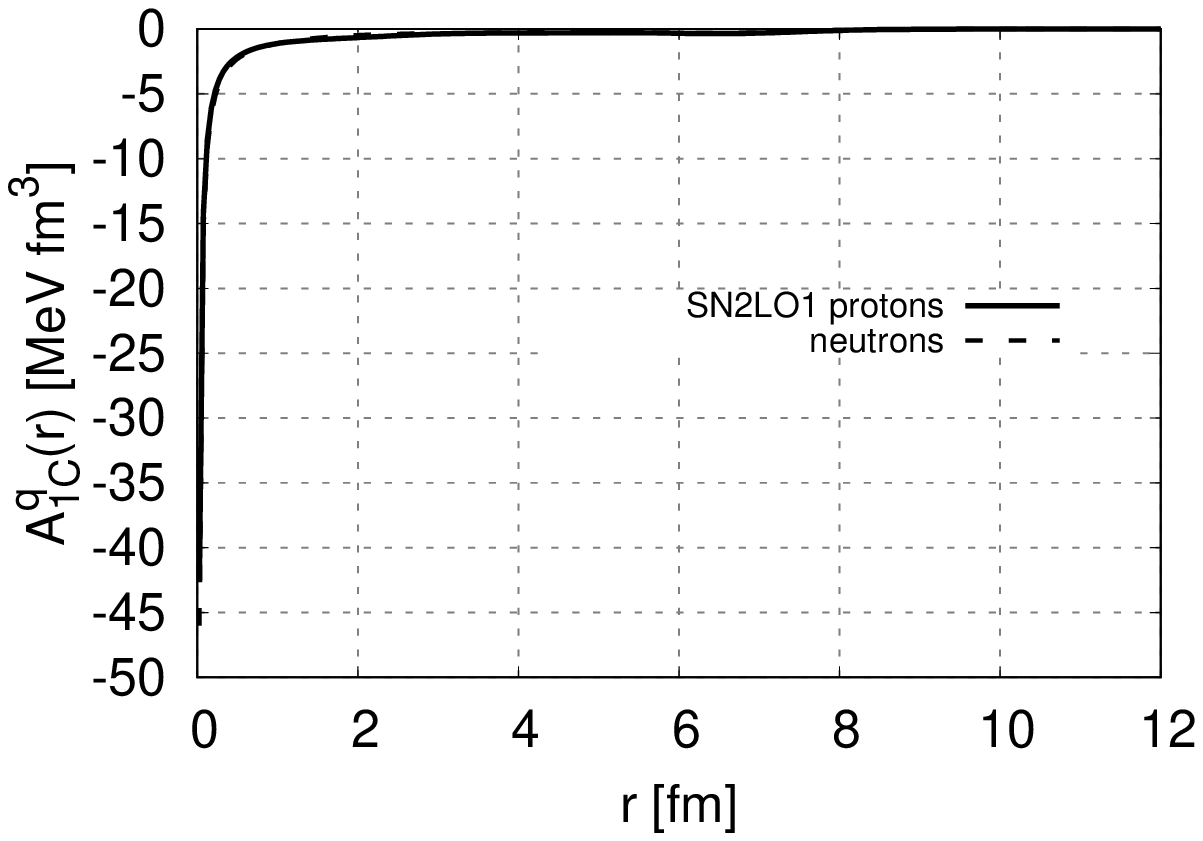}\\
\includegraphics[width=0.45\textwidth,angle=0]{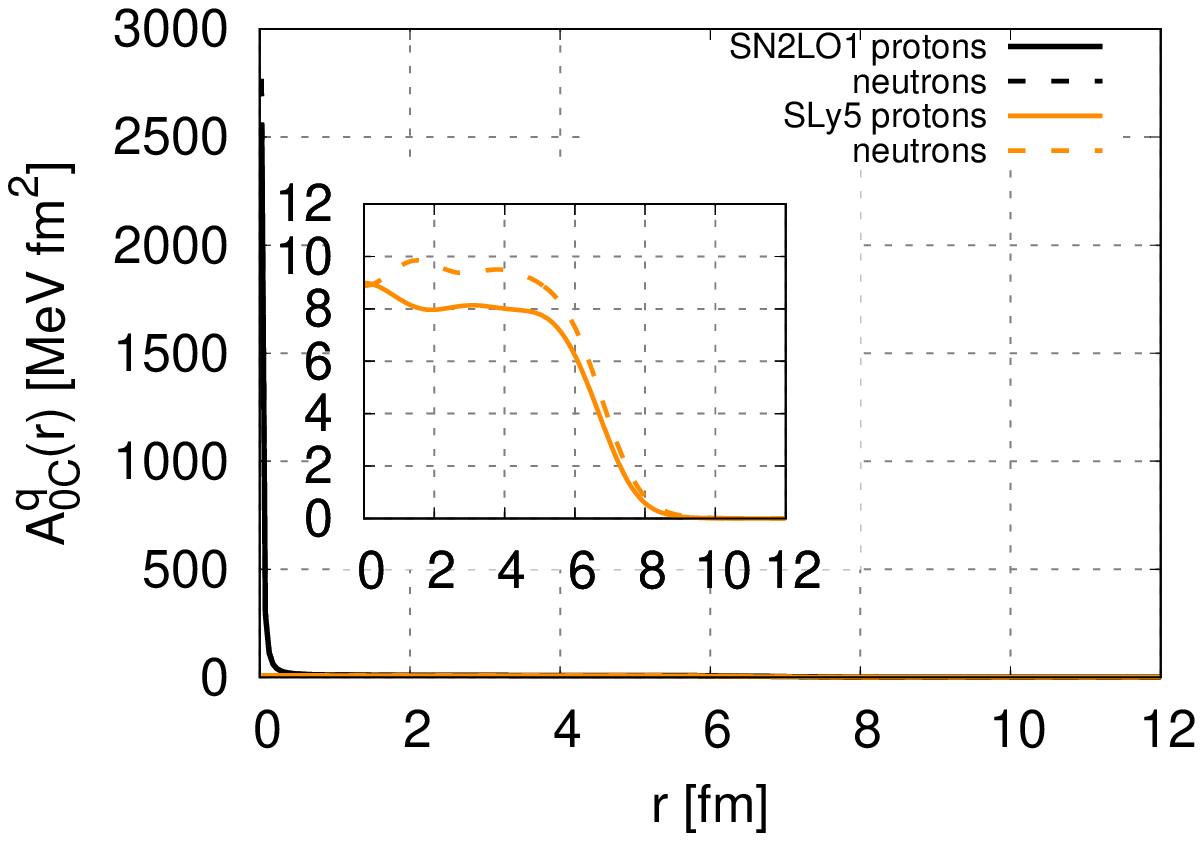}
\includegraphics[width=0.45\textwidth,angle=0]{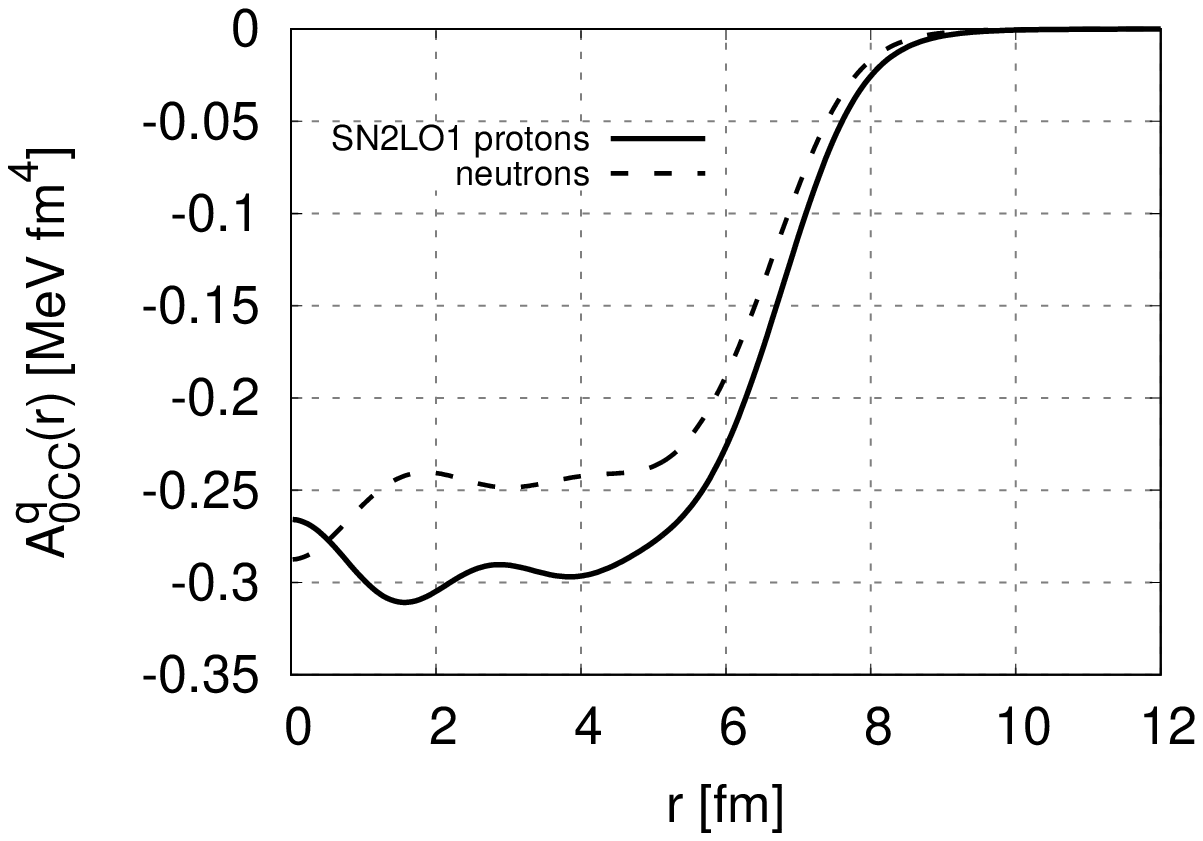}
\end{center}
\caption{(Colors online) Same as Fig.~\ref{208pb:field}, but for centrifugal fields given in Eq.~(\ref{sp:eq:4th}).}
\label{208pb:fieldc}
\end{figure}

\begin{figure}[!h]
\begin{center}
\includegraphics[width=0.45\textwidth,angle=0]{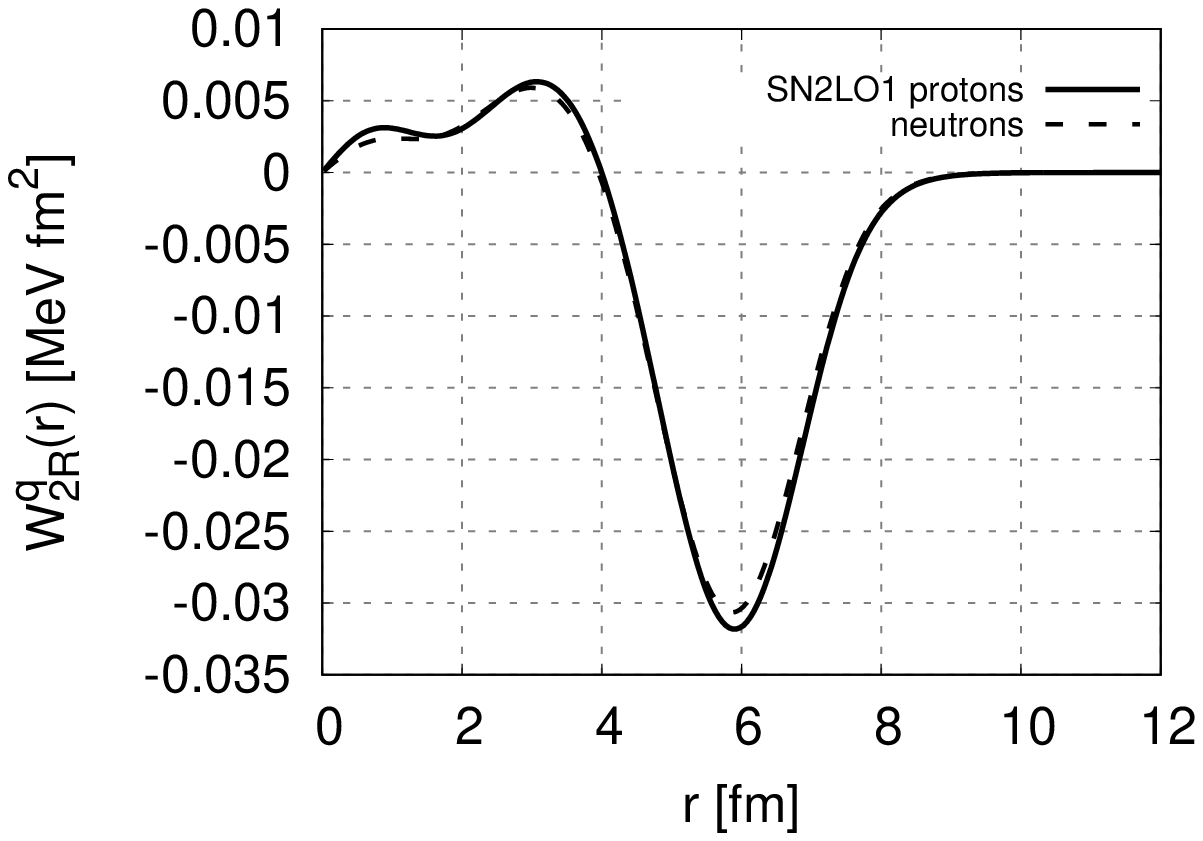}
\includegraphics[width=0.45\textwidth,angle=0]{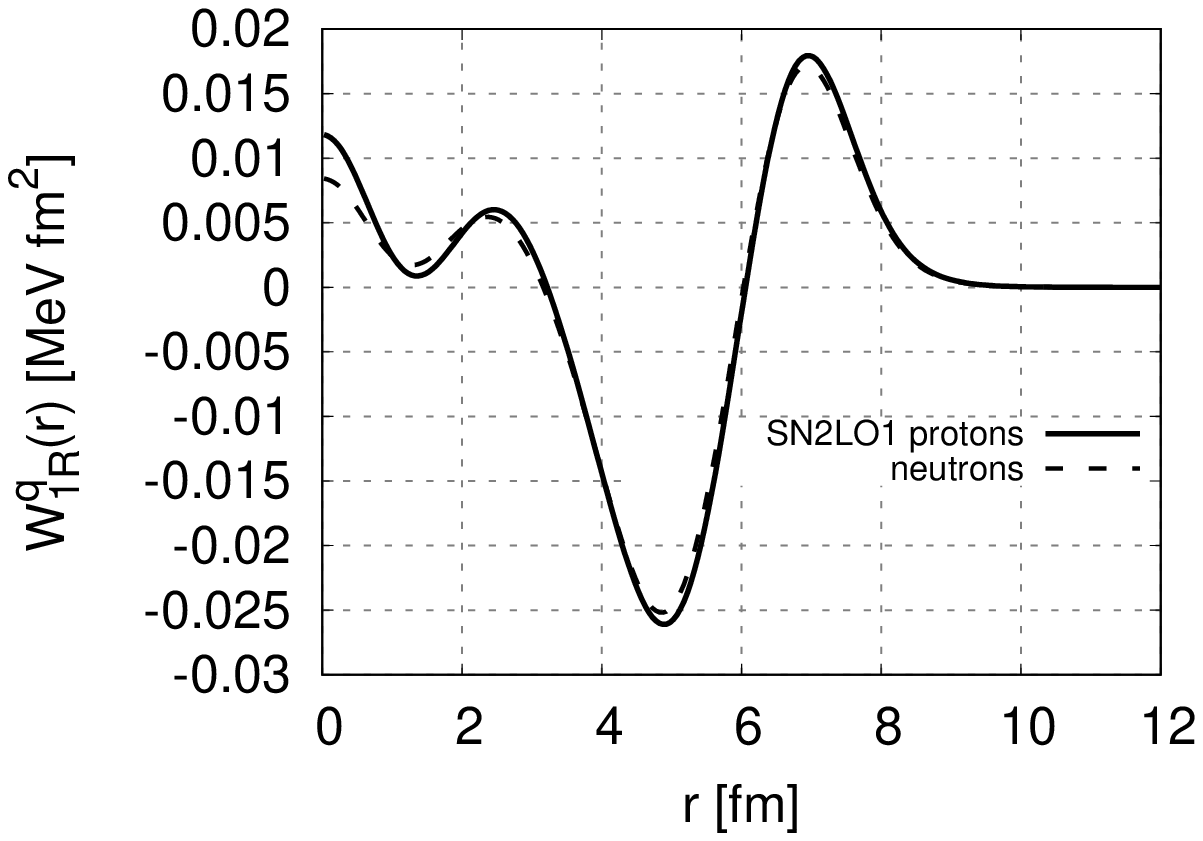}
\includegraphics[width=0.45\textwidth,angle=0]{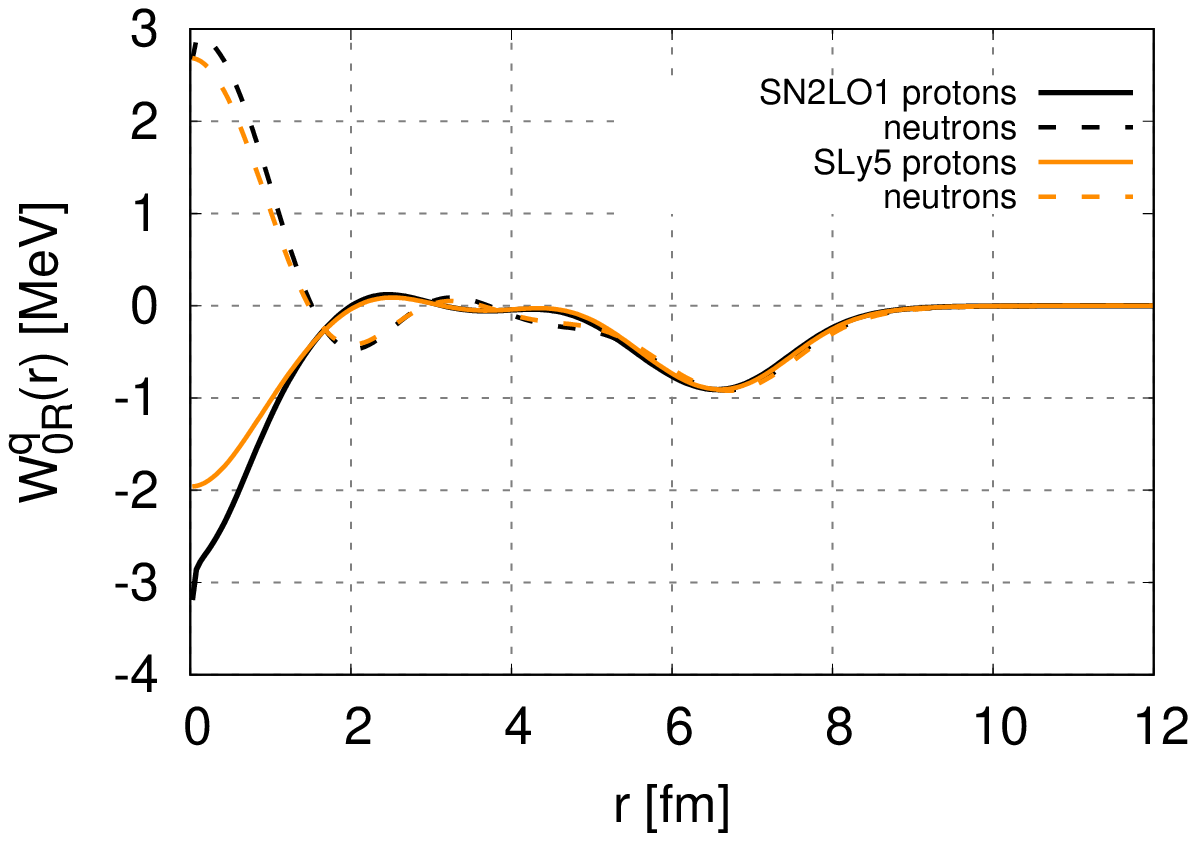}
\includegraphics[width=0.45\textwidth,angle=0]{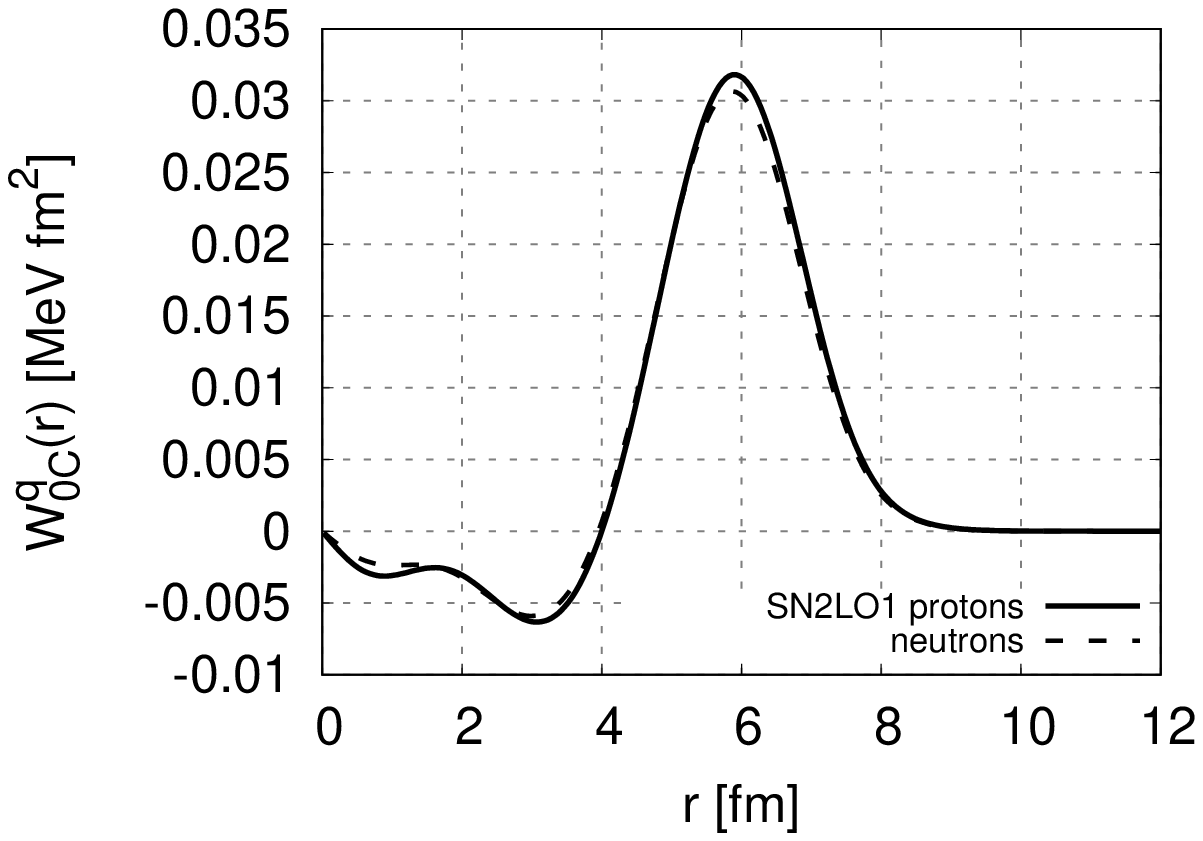}
\end{center}
\caption{(Colors online) Same as Fig.~\ref{208pb:field} but for the spin-orbit fields given in Eq.~(\ref{sp:eq:4th}).}
\label{208pb:so}
\end{figure}

The coefficients in Eq.~(\ref{sp:eq:4th}) are defined as
\begin{eqnarray}
 A^q_4 &=& \,  C^{ M \rho}_-  \, \rho_0  \, + 2 \, \, C^{ M \rho}_1 \, \rho_q \label{eq:a4} \;,\\ 
 A^q_3 &=&  2 \,  C^{M \rho}_-  \, \rho_0^{(1)}   +  4 \, C^{ M \rho}_1 \, \rho_q^{(1)}\;, \label{eq:a3}\\
A^q_{2 R}  &=&  -  \frac{\hbar^2}{2m}  -   C_-^{\tau} \ \rho_0  -  2 C_1^{\tau} \rho_q  +  C_-^{ M \rho} \left[  3 \rho_0^{(2)} - 6 \tau_{R,0} - 2  \tau_{C,0}   \right] +   2 \, C_1^{ M \rho}\left[ 3 \rho_q^{(2)} - 6 \tau_{R,q} - 2  \tau_{C,q}   \right]\;, \label{eq:a2r}\\
A^q_{2 C}   &=& - 2 \,  C^{ M \rho}_-  \, \rho_0  \, - \, 4 \, C^{ M \rho}_1 \, \rho_q \;,\label{eq:a2c}  \\
A^q_{1 R} &=&  - \ C_-^{\tau} \rho_0^{(1)}  -  2 C_1^{\tau} \rho_q^{(1)}  + 2 \,  C_-^{ M \rho} \left[\rho_0^{(3)}-3 \tau_{R,0}^{(1)} -  \tau_{C,0}^{(1)}   \ \right] + 4 \, C_1^{ M \rho}\left[\rho_q^{(3)}-3 \tau_{R,q}^{(1)} -  \tau_{C,q}^{(1)}   \ \right] \;,  \label{eq:a1r} \\
A^q_{1 C} &=&   2 \,  C^{ M \rho}_-  \, \left(-\rho_0^{(1)} + 2\frac{\rho_0}{r}\right)  \, + \, 4 \, C^{ M \rho}_1 \, \left(-\rho_q^{(1)} + 2\frac{\rho_q}{r}\right) \;, \label{eq:a1c}  \\ 
A^q_{0 R} &=&    U_q (r) +   C_-^\tau \frac{\rho_0^{(1)}}{r} + 2 \ C_1^\tau \frac{\rho_q^{(1)}}{r} \;, \nn \\
           &+&  2 \,  C^{M \rho}_- \left[ 3 \frac{\tau_{R,0}^{(1)}}{r} + \frac{\tau_{C,0}^{(1)}}{r} - \frac{\rho_0^{(3)}}{r} \right] + 4 \,  C^{ M \rho}_1 \left[ 3 \frac{\tau_{R,q}^{(1)}}{r} + \frac{\tau_{C,q}^{(1)}}{r} - \frac{\rho_q^{(3)}}{r} \right]  \;, \label{eq:a0r}\\
A^q_{0 C} &=&   \frac{\hbar^2}{2m} + C_-^\tau \rho_0 + 2 C_1^\tau \rho_q +  C^{M \rho}_- \left[ 2 \, \tau_{R,0} + 4 \tau_{C,0} + 2 \frac{\rho_0^{(1)}}{r} -  \rho_0^{(2)}  - 6 \frac{\rho_0}{r^2} \right] \nn \label{eq:a0c} \\
           &+& 2 \,  C^{ M \rho}_1 \left[ 2 \, \tau_{R,q} + 4 \tau_{C,q} + 2 \frac{\rho_q^{(1)}}{r} -  \rho_q^{(2)}  - 6 \frac{\rho_q}{r^2} \right] \;, \\
A^q_{0 CC} &=&  \, C^{ M \rho}_-  \rho_0 + 2 \, C^{ M \rho}_1  \rho_q \label{eq:a00c} \,. 
\end{eqnarray}
Here we used the shorthand notation $C_{-}^x=C_0^x-C_1^x$ with $x=\rho,\Delta\rho,\dots$. The exponent ($i=1,2,3,4$) in the densities stands for the derivative order. Finally, the central field appearing in the previous equation reads
\bqr
\label{eq:uqD}
U_q (r)&=& 2  C^{\rho}_- \rho_0 + 4 \, C^{\rho}_1 \rho_q + 2 C^{\Delta \rho}_- \Delta \rho_0 + 4 \, C^{\Delta \rho}_1  \Delta \rho_q +  C^{\tau}_- \tau_0 + 2 \, C^{\tau}_1  \tau_q \nn \\
&+&  2  C^{  (\Delta \rho)^2}_- \Delta \Delta \rho_0 + 4 \, C^{ (\Delta \rho)^2}_1  \Delta \Delta \rho_q C^{M \rho}_- \left[Q_0 -  2 \nabla_\mu \nabla_\nu \tau_{\mu \nu , 0} \right] + 2 C^{M \rho}_1 \left[ Q_q - 2  \ \nabla_\mu \nabla_\nu\tau_{\mu\nu , q}\right]  \nn \\
&+&  C^{\nabla J}_- \nabla \cdot J_0 + 2 C^{\nabla J}_1 \nabla \cdot J_q.
\eqr
This field is obtained through the variational principle varying the matter density $\rho$, and  it receives contributions from both N1LO and N2LO terms. In Fig.~\ref{208pb:field} we show the coefficients $A^q_R$ and the central field $U_q$ obtained with a fully converged HF calculation (cf Tab.~\ref{tab:inter}) in $^{208}$Pb using a N2LO pseudo-potential. We refer the reader to Sec.~\ref{sec:fit} for more details on this parametrisation. 
On the same figure we also report the corresponding values obtained with SLy5. As it should be, SLy5 induces non-zero contributions only for the terms originating from the N1LO part of the functional. In Fig.~\ref{208pb:fieldc} we show the other set of fields appearing in Eq.~(\ref{sp:eq:4th}) and corresponding to the centrifugal parts. These fields are active only for non-zero orbital momentum states. All the fields behave normally around $r=0$ apart from the $A^q_{1c},A^q_{0c}$ that present a divergency. Such a behaviour, which already exists at N1LO level for the centrifugal field, is actually not a problem as we will see in Sec.~\ref{sec:asym} when we examine the asymptotic properties of our 4th order differential equation. We will then demonstrate that there exists a particular solution of Eq.~(\ref{sp:eq:4th}) that exhibits no divergency.
Although we have only one explicit spin-orbit term in the effective interaction, we obtain four distinct contributions to the mean-field equation
\begin{eqnarray}
  W^q_{0 R } (r)  &= & - C_\alpha \left[ C^{T}_- \frac{J_0}{r} + 2 C^{T}_1\frac{J_q}{r} + C^{\nabla J}_- \frac{\rho_0^{(1)}}{r} + 2 C^{\nabla J}_1 \frac{\rho_q^{(1)}}{r} \right] \label{eq:so1}  \\
&& +   C_\alpha \left[2 C^{(4) M s}_- \left( \frac{J_0}{r^3} -\frac{J_0^{(1)}}{r^2} - \frac{V_0(r)}{r} + 2 \frac{K_0(r)}{r}\right) + 4 C^{(4) M s}_1 \left( \frac{J_q}{r^3} -\frac{J_q^{(1)}}{r^2} - \frac{V_q(r)}{r}+ 2 \frac{K_q(r)}{r}\right) \right]\;, \nn \\
W^q_{0 C } (r) &=& C_\alpha  \left[ - 2 C^{ M s}_- \frac{ J_0 (r)}{r} - 4  C^{ M s}_1  \frac{J_q (r)}{r}  \right],\label{eq:so2}\\
 W^q_{1 R} (r) &=& C_\alpha     \left[ 2 C^{ M s}_- \left( \frac{J_0^{(1)}(r)}{r} - \frac{ J_0 (r)}{r^2} \right) + 4  C^{ M s}_1 \left(\frac{ J_q^{(1)}(r)}{r} - \frac{ J_q (r)}{r^2} \right) \right],\label{eq:so3}\\
 W^q_{2 R} (r) &=&  C_\alpha  \left[ 2 C^{ M s}_- \frac{ J_0 (r)}{r}+ 4  C^{M s}_1  \frac{ J_q (r)}{r}  \right].\label{eq:so4}
\end{eqnarray}
This is a very interesting feature of our functional which appears to have more flexibility than N1LO. This new dependence could be of particular interest in different situations, by instance in adjusting centroids of single particle states without the need of using an explicit tensor term. Moreover, these terms are associated with the first two derivatives in the differential equation, contrary to the standard Skyrme interaction, and one of them is a centrifugal term. Such a term could thus allow to act on the single-particle levels with a new dependency in $l$. It is worth mentioning that several Skyrme functionals use different coupling constants in the spin-orbit sector to enrich the freedom of the corresponding field~\cite{rei99}. In such a case, the link with the underlying interaction is then broken. The new N2LO functional presented here has the advantage of keeping such a link and also gaining a more complex spin-orbit structure, thus making it a suitable candidate for multi-reference calculations. In Fig.~\ref{208pb:so}, we show the different spin-orbit contributions. The current parametrisation SN2LO1 leads to relative small values, but we should not exclude a priori the possibility of finding significative corrections with a different set of parameters.

\subsection{Asymptotic properties}\label{sec:asym}

Before entering the numerical details of the solution of Eq.~(\ref{eq:hf}), we want to prove that a solution with a well-behaved asymptotic behaviour (origin and infinity) exists. It has been well established for the standard Skyrme second-order differential equation~\cite{vau72} that the radial part of the wave-function Eq.~(\ref{fosphe}) behaves as $R_\alpha \propto r^{l+1}$ at the origin so that it compensates the behavior of the centrifugal term which diverges as $1/r^2$. In the case of the present fourth-order differential equation, this result is a priori no longer true. We thus assume that $R_\alpha(r) \propto r^\beta$ around  $r=0$ and determine the possible physical value for $\beta$. We insert it in HF equations given in Eq.~(\ref{eq:hf}) and we obtain
\begin{eqnarray}
\epsilon_\alpha r^4 = && \beta (\beta-1) (\beta-2) (\beta-3) A_4 + \beta (\beta-1) (\beta-2) A_3 r +  \beta (\beta-1) A_{2 R} r^2 +  \beta A_{1 R} r^3  \nn \\
                                   && + A_{0 R} r^4 + \ell (\ell+1) \left[  \beta (\beta-1) A_{2 C} +  \beta A_{1 C} r +  A_{0 C} r^2 + \ell (\ell+1) A_{0 CC} \right] \nn \\
                                  && + \left( j(j+1) - l(l+1) - \frac{3}{4} \right) \left[ W_{0 R} r^4 + \ell (\ell+1) W_{0 C} r^2 + \beta W_{1 R} r^3 + \beta (\beta -1) W_{2 R} r^2 \right]\;.
\end{eqnarray}
All non relevant single-particle quantum numbers are omitted in this discussion to make the notation lighter.
By inspecting the formal expressions of the coefficients $A_i$ in Eqs.~(\ref{eq:a4}-\ref{eq:a00c}), we observe that some fields diverge around origin
\begin{eqnarray}
A_{1C}\xrightarrow[r\rightarrow0]{} \frac{1}{r}\;,\\
A_{0C}\xrightarrow[r\rightarrow0]{} \frac{1}{r^2}\;.
\end{eqnarray}
The term $A_{0R}$ does not diverge since the derivative of the density is zero at the origin. This is typically the case of nuclear densities, even in the case of strong shell effects~\cite{ber01}. The spin-orbit fields have no divergence, so we can drop them. To have a well-behaved wave-function at $r=0$ we thus need to check that only the following terms give zero
\begin{eqnarray}
\beta (\beta-1) (\beta-2) (\beta-3) A_4 + \beta (\beta-1) (\beta-2) A_3 r +  \beta (\beta-1) A_{2 R} r^2 +  \beta A_{1 R} r^3  \nn \\
                                   + A_{0 R} r^4 + \ell (\ell+1) \left[  \beta (\beta-1) A_{2 C} +  \beta A_{1 C} r +  A_{0 C} r^2 + \ell (\ell+1) A_{0 CC} \right] \approx 0\;.
\end{eqnarray}
First we  notice that $A_{3},A_{2R}$ and $A_{1R}$ do not diverge at the origin. When multiplied by powers of $r$, they thus go to zero at the origin.
By inspecting Eqs.~(\ref{eq:a4}-\ref{eq:a00c}), we can then notice that to leading order the following relations hold
\begin{eqnarray}
\ \ \ \ \ \   A_{2 C} = - 2 A_4 \ \ \ \ \ \ A_{1 C} = 4 A_4  \ \ \ \ \ \   A_{0 C} = - 6 A_4 \ \ \ \ \ \  A_{0 CC} = A_4\;,
\end{eqnarray}
so that we can simplify 
\begin{equation}
\beta (\beta-1) (\beta-2) (\beta-3) A_4 + \ell (\ell+1) \left[  \beta (\beta-1) A_{2 C} +  \beta A_{1 C} +  A_{0 C} + \ell (\ell+1) A_{0 CC} \right] \simeq 0\;.
\end{equation}
We finally obtain
\begin{equation}
\beta^ 4 - 6 \beta^3 + \beta^2 \left( - 2 \ell^2 - 2 \ell +11 \right) + 6 \beta \left(\ell^2 + \ell -1 \right) + \ell (\ell+1) \left( \ell^2 + \ell - 6 \right) \simeq 0\;.
\end{equation}
This equation has 4 solutions
\begin{eqnarray}
 \beta = 2 - \ell ,\ \ \ \ \ \ \beta = - \ell , \ \ \ \ \ \   \beta =\ell + 1 , \ \ \ \ \ \ \beta  =\ell + 3\,.
\end{eqnarray}
The first two solutions diverge for some specific values of $\ell$ and can not represent the physical behaviour of the radial wave function. The last two solutions are physically well-behaved but since the nuclear density needs to be non-zero at the center of the nucleus, only the solution $\beta =\ell + 1 $ can be accepted. The radial part has therefore the same behaviour for N1LO and N2LO. At infinity, all the fields vanish as one can easily see from Figs.\ref{208pb:field}-\ref{208pb:so}, thus we can recover the typical asymptotic behaviour of the solutions of the  N1LO functional.


\subsection{Numerical methods to solve 4th order equations}

The solution of HF equations with 4th order derivative terms represent a major numerical challenge. The standard technique for N1LO is usually to project the HF equations on an Harmonic Oscillator basis, since one can use particular properties of orthonormal polynomials to avoid the explicit numerical derivation~\cite{hosphe}. However, the main inconvenient is the slow convergence as a function of the number of basis states, as compared to the solution of the HF equations via direct integration~\cite{sch15}. We have thus decided to develop  a new  numerical solver named \verb|WHISKY|~\cite{bec17}: the code has been built in a modular way so it can accept the central part of the N$\ell$LO Skyrme pseudo-potential with $\ell=1,2,3$. The code has been written aiming at using it into a fitting procedure. Therefore it has been conceived to be fast and accurate. To conciliate high accuracy and reduced execution time, we have decided to use a two-basis method to solve HF equations~\cite{sch12}. The 4th order differential equation governing the properties of single-particle states is then solved using the finite-difference method and more particularly the Hooverman method~\cite{hoo72}. With this method, we obtain a wave-function for each point of the mesh for each $(\ell,j,q)$-block. As a consequence the number of basis functions grows quite quickly, especially when we include pairing correlations (see Sec.~\ref{subsec:pair}) so that we introduced an auxiliary Wood-Saxon (WS) basis and an additional energy cut-off. Since the WS wave-functions are reasonably close to the final single-particle solutions, the number of basis states to ensure convergence is quite reduced. An alternative to the WS basis would be the use of the self-consistent HF basis. However, we did not explore this possibility: since we are not currently working with very neutron rich nuclei, a WS approximation is expected to give a result close to the final solution. We plan to add this option to explore the properties of the extended N2LO functional close to stability in the next version of the code.
\begin{figure}[!h]
\begin{center}
\includegraphics[width=0.45\textwidth,angle=0]{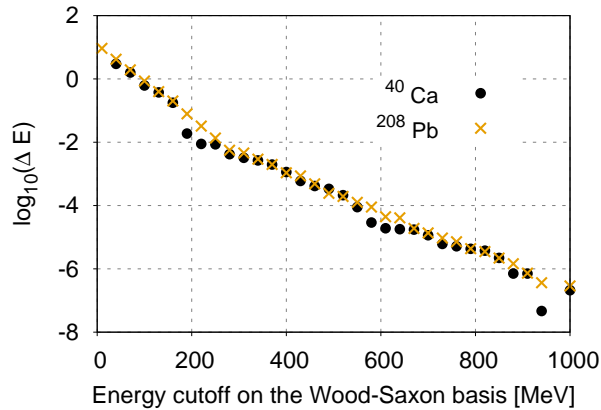}
\end{center}
\caption{(Colors online) Precision obtained with WHISKY against LENTEUR as a function of the cutoff energy in the Wood-Saxon basis for $^{40}$Ca (+) and $^{208}$Pb ($\times$). See text for details.}
\label{208pb:prec2}
\end{figure}

In Fig.~\ref{208pb:prec2}, we compare the accuracy of our HF code against the HF code named \verb|LENTEUR|~\cite{rot09a,rot09b} as a function of the intermediate WS basis size. The calculations are done in both cases using SLy5 interaction~\cite{cha97} with Coulomb included and a mesh of $h=0.05$ fm  within  a box of 20 fm.
 It is worth reminding that the code  \verb|LENTEUR| works with a similar two-basis method: HF  and r-space representation with direct integration of HF equation in coordinate space~\cite{ben05}. The total energy difference for different nuclei obtained with the two codes  is defined as $\Delta E=\left|E_{\text{WHISKY}}-E_{\text{LENTEUR}}\right|$. We observe that the accuracy of our code is very good in a reasonably small basis size. By considering states up to 300 MeV we obtain an accuracy of $\approx1$ keV and an execution time of a few seconds. In Tab.~\ref{tab:energy}, we give a more detailed comparison of the resulting energies for a fully converged calculation in $^{208}$Pb using \verb|LENTEUR|  and \verb|WHISKY| with a cut-off of 300 MeV in WS basis. We see that the agreement is very good (8 keV at worst). We conclude that the basis size we have chosen is clearly an excellent compromise of efficiency and accuracy since all energy contributions are described by the two codes at the keV level of accuracy. This cutoff is consequently used in the fit.
\begin{table}
\begin{center}
\begin{tabular}{c|cc}
\hline
\hline
 \multicolumn{3}{c}{$^{208}$Pb}\\
\hline 
[MeV] &  \verb|WHISKY|  &  \verb|LENTEUR|  \\
\hline
Total Energy  &-1636.10{\textbf 6} & -1636.10{\textbf 5}\\
Kinetic energy  & 3874.7\textbf{89} & 3874.7\textbf{95}\\
Field energy  & -6209.6\textbf {42} & -6209.6\textbf {50}\\
Spin-orbit       & -99.081 & -99.081 \\
Direct Coulomb & 829.143 & 829.143\\
Exchange Coulomb & -31.314 &-31.314\\
\hline
\hline
\end{tabular}
\end{center}
\caption{Energies obtained by WHISKY and LENTEUR with self-consistent HF calculations using the SLy5 interaction. The differences appear on the last digits and are written in bold.
}
\label{tab:energy}
\end{table}

The code  \verb|LENTEUR|  accepts only N1LO Skyrme-like functionals. Therefore, in order to test the energy contribution of the terms originated from higher order derivatives, we benchmarked our code against the latest version of \verb|MOCCA|~\cite{rys15,mocca}. \verb|MOCCA| is a 3D solver working in a cubic box and using imaginary-time algorithm to solve the HF equations~\cite{bon05}. For the current comparison, we used a mesh of $dx=0.4$  fm  and 32 points in each direction. Since we deal with spherical even-even nuclei, we can impose several symmetries and thus perform the calculations only in  one octant of the whole box. See Ref.~\cite{rys15} for more details. 
For our tests we have used SLy5 and added a random set of higher order parameters. The results are presented in Tab.~\ref{tab:energy2}. The different energy terms refer to the different components of the N2LO functional as given in Eq.~(\ref{eq:ef:DKo}). In this case the total energy difference between the two codes is at the level of 10 keV on the total energy. This is also the typical discrepancy between the different 4th order terms of the N2LO functional. This is a very strong test since the two codes have been developed in a completely independent way and moreover they use completely different algorithms to solve the HF equations.
\begin{table}
\begin{center}
\begin{tabular}{c|cc}
\hline
\hline
 \multicolumn{3}{c}{$^{208}$Pb}\\
\hline 
[MeV] &  \verb|WHISKY|  &  \verb|MOCCA|  \\
\hline
Total Energy  & -1539.2{\textbf {53}} & -1539.2{\textbf {63}}\\
Total energy N2LO  & 89.{\textbf {278}} & 89.{\textbf {360}}\\
E[$(\Delta\rho)^2$] & 4.39{\textbf 4} & 4.39{\textbf 5}\\
E[$\rho Q$]       & 37.4{\textbf {77}} & 37.4{\textbf {88}}\\
E[$\tau^2$]       & 27.2{\textbf {12}} & 27.2{\textbf {21}}\\
E[$\tau_{\mu\nu}\tau_{\mu\nu}-\tau_{\mu\nu}\nabla_{\mu}\nabla_{\nu}\rho$]       & 19.8{\textbf {55}} & 19.8{\textbf {61}}\\
E[$K_{\mu\nu\kappa}K_{\mu\nu\kappa}$] & 0.0546{\textbf 0} & 0.0546{\textbf 1}\\
E[$J_{\mu\nu}V_{\mu\nu}$] & 0.3385{\textbf 0} & 0.3385{\textbf 8}\\
\hline
\hline
\end{tabular}
\end{center}
\caption{Comparison of the results for WHISKY and MOCCA: different N2LO functional contributions to the total energy after a self-consistent calculation with a toy N2LO interaction. The discrepancies are presented in bold.
}
\label{tab:energy2}
\end{table}

\subsection{Pairing correlations}\label{subsec:pair}

Once we move away from closed-shell nuclei, we need to consider extra pairing correlations~\cite{bri05}. To this purpose, we have generalized the \verb|WHISKY| code to solve the complete HFB equations. Since we use a two-basis method, we first solve the HF equations in coordinate space and then we transform back to the WS basis.
The HFB equations in this basis  read~\cite{pas08}
\begin{eqnarray}
  \sum_{\alpha'}(h_{\alpha'\alpha}^{lj,q}- \mu_{F}^{q})U^{nlj,q}_{\alpha'}+\sum_{\alpha'}\Delta_{\alpha \alpha '}^{lj,q}V^{nlj,q}_{\alpha'}&=&E^{nlj,q}U^{nlj,q}_{\alpha} ,
      \\
  \sum_{\alpha'}\Delta^{lj,q}_{\alpha \alpha'}U^{nlj,q}_{\alpha'} -\sum_{\alpha'}(h^{lj,q}_{\alpha'\alpha}- \mu_{F}^{q})V^{nlj,q}_{\alpha'} &=&E^{nlj,q}V^{nlj,q}_{\alpha} ,
\label{paper:eq:HFBeq}
\end{eqnarray}
where $\mu_{F}^{q}$ is the chemical potential and $U^{nlj,q}_{\alpha}$ and $V^{nlj,q}_{\alpha}$ are the Bogoliubov amplitudes for the quasiparticle of energy $E^{nlj,q}$, $\alpha$ is the index of the WS basis and $n$ is the index of the quasi-particle state. The field $h_{\alpha'\alpha}^{lj,q}$ is derived from Eq.~(\ref{sp:eq:4th}) {\it via} a unitary transformation.

For the pairing channel we used a simple pairing interaction of the form~\cite{ber91,gar99}
\begin{eqnarray}
\label{pairing_int_contact}
\qquad \quad v(\mathbf{r}_{1},\mathbf{r}_{2})=V^q_{0}\left[ 1- \eta \left( \frac{\rho_0\left(
\mathbf{R}\right)}{\rho_{sat}}\right)^{}\right]
\delta(\mathbf{r}), 
\end{eqnarray} 
where $\mathbf{R}=(\mathbf{r_1}+\mathbf{r_2})/2$ is the center of mass of the two
interacting particles and $\mathbf{r}=\mathbf{r_1}-\mathbf{r_2}$ is their mutual distance. In the present article we use the so-called volume shape~\cite{san05} with parameter  $V^n_{0}=V^p_{0}=-200$ MeV.fm$^3$, $\eta=0$, and $\rho_{sat}=0.16$ fm${^{-3}}$. Since this interaction has an ultraviolet divergency~\cite{bul02}, we use a simple cut-off procedure in quasi-particle space $E_{cut}=60$ MeV. For more details on this topic we refer to Ref.~\cite{bor06}. The choice of the pairing interaction is crucial to determine properties of nuclei far from stability~\cite{dob94,pas13b}. At present we followed the Saclay-Lyon fitting protocol, so we decoupled the problem in two steps. After the complete fit of the N2LO functional, the $V_{0}$ parameters can be fixed to pairing effects. In this article we have used prefixed values of the $V_0$ parameters, but we plan to extend our fitting procedure to take into account also pairing effects more precisely~\cite{gom92}. The pairing interaction for protons and neutrons is not necessary the same, since Coulomb effects should also taken into account in the calculation of proton Cooper pairs~\cite{nak11}. We plan to include such effects in the next version of the code.

In Tab.\ref{tab:energy:pair}, we compare \verb|WHISKY| against \verb|LENTEUR| for $^{120}$Sn and SLy5 interaction plus volume pairing Eq.\ref{pairing_int_contact}.
We observe that the accuracy is remarkably high. The small discrepancy of 4 keV originates from a different definition of cut-off in single particle states: \verb|LENTEUR|  operates with a cut-off on the total angular momentum $j$ of the quasi-particle states entering the calculation, while \verb|WHISKY| operates with a cut-off on the orbital angular momentum.

{\begin{table}
\begin{center}
\begin{tabular}{c|cc}
\hline
\hline
 \multicolumn{3}{c}{$^{120}$Sn}\\
\hline 
[MeV] &  \verb|WHISKY|  &  \verb|LENTEUR|  \\
\hline
   Total energy  &  -1018.81\textbf{4}  &  -1018.81\textbf{8}    \\  
     Kinetic energy  &  2188.1\textbf{27}  & 2188.1\textbf{42}  \\  
      Field energy  & -3485.1\textbf{15}  &     -3485.1\textbf{31}   \\  
      Spin-orbit energy  &   -55.00\textbf{0}  &   -55.00\textbf{1}   \\  
      Coulomb (direct)  &  367.336  &  367.336    \\  
       Coulomb (exchange)  & -19.147 &  -19.147    \\ 
       Neutron pairing energy  &  -15.01\textbf{4}  &    -15.01\textbf{7}   \\
\hline
\hline
\end{tabular}
\end{center}
\caption{Comparaison between the energies obtained by WHISKY and lenteur with self-consistent HFB calculations using the SLy5 interaction. See text for details }
\label{tab:energy:pair}
\end{table}

In Fig.\ref{pair}, we compare the total density $\rho(r)$ for $^{120}$Sn obtained with the two codes and also the pairing density $\tilde{\rho}$. Following Refs.~\cite{dob84,ben05} we define it as

\begin{eqnarray}
\tilde{\rho}_{q} (r)               & = & -\sum_{nlj} \  \frac{(2 j + 1 )}{4 \pi} \ \frac{V^{nlj,q}(r)U^{nlj,q}(r)}{r^2} \,, 
\end{eqnarray}
where $V^{nlj,q}(r), U^{nlj,q}(r)$ are the quasi-particle amplitudes expressed in r-space.
\begin{figure}[!h]
\begin{center}
\includegraphics[width=0.45\textwidth,angle=0]{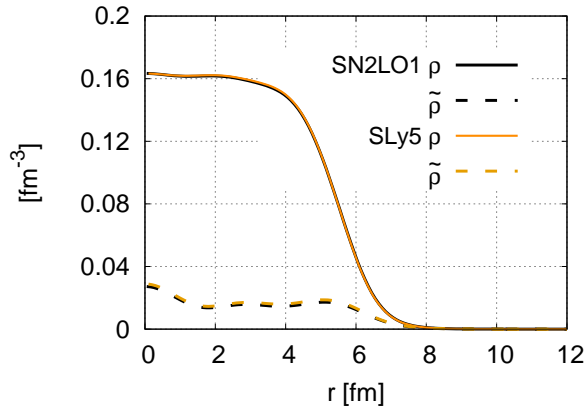}
\end{center}
\caption{(Colors online) Isoscalar particle density and pairing density for $^{120}$Sn obtained with a self-consistent mean-field calculation with the SLy5 interaction.}
\label{pair}
\end{figure}
The agreement is excellent, thus demonstrating the very high accuracy of our new NEDF solver.

\section{Fit of N2LO interaction}\label{sec:fit}

To fit the N2LO pseudo-potential we adopted a modified version of the Saclay-Lyon fitting protocol~\cite{cha97,kou12}: the protocol includes here both properties of some selected double-magic nuclei  and some basic properties of the infinite nuclear medium as saturation density, incompressibility and the equation of state of pure neutron matter (PNM) derived from realistic nucleon-nucleon interactions~\cite{Wiringa}. We consider \emph{all} terms of the interaction, and we treat spurious center of mass motion with the usual one-body approximation~\cite{cha97,ben03}. We also assume equal neutron and proton masses and we use the value $\frac{\hbar^2}{2m}=20.73553$ MeV.fm$^2$~\cite{cha97}.

\subsection{Fitting protocol}

To obtain the parameters of the pseudo-potential we need to minimize the following penalty function~\cite{dob14}
\begin{eqnarray}\label{eq:chi2}
\chi^2=\sum_{i=1}^M\frac{\left( \mathcal{O}_i -f_i(\mathbf{p}) \right)^2}{\Delta \mathcal{O}_i^2}\;,
\end{eqnarray}
where the sum runs over all the $M$ (pseudo)-observables $\mathcal{O}_i $ we want to constraint in our fit, $f_i $ is the value obtained with our solver for a given array of parameters $\mathbf{p}=\left\{ t_0,t_1,t_2,\dots \right\}$, while $\Delta \mathcal{O}_i$ is the weight we give to each point in the fit. Let's mention that $\Delta \mathcal{O}_i$ does not correspond necessarily to the experimental uncertainty. In Tab.~\ref{Totcont}, we give the actual constraints we used to build the $\chi^2$ function in Eq.~(\ref{eq:chi2}). On top of this constraints, we paid particular attention in tuning the spin-orbit parameter $W_0$ to some specific range of acceptable values. Finally, it is worth noticing that during the $\chi^2$ minimisation the parameters $\mathbf{p}$  cannot vary freely: in order to avoid finite-size instabilities~\cite{hel13},  the critical densities in all channels are computed at each iteration, and an asymmetric constraint is imposed in terms of a penalty function
\begin{eqnarray}\label{eq:chi2:spurious}
\chi^2_{fs}=\sum_{\alpha}\exp^{-2\beta \left(\mathcal{O}_{\alpha} -\rho_{crit}\right)}\;,
\end{eqnarray} 
where $\mathcal{O}_{\alpha=(S,M,T)}$ is the lowest density at which an instability appears in symmetric nuclear matter (SNM). $\rho_{crit}$ is an \emph{empirical} value defined in Refs.~\cite{hel13,Pas15T} to avoid unphysical instabilities. $\beta$ is an arbitrary parameter ($\beta =10$ here) fixed in such a way that the penalty function grows very fast when we approach the critical density from below, but gives no contribution when above it. This constraint is applied in all  channels for which we calculate the response function of the system (see Sec.~\ref{sec:finitesize}). Finite-size instabilities may also have important impact at high density on astrophysical applications such as the neutrino mean free path~\cite{pas14bsk}. However, in this work, we concentrate ourselves on finite-size instabilities only in densities ranges that are relevant for finite nuclei. In other words, we allow in this preliminary work the appearance of instabilities at densities above $\rho_{crit}$ which is slightly above saturation density.

\begin{table}
\centering
\begin{tabular}{|l| c  c c c |}
  \hline
   {\centering \large  \quad  \ \ Fit Constraints}   &  $\mathcal{O}_i$ & \ $\Delta \mathcal{O}_i$  & Units & Reference \\
   \hline
   \hline
\ \  \textbf{Infinite nuclear matter} \ \ & & & & \\
  \ $\rho_{sat}$  & 0.1600  & 0.001  & fm$^{-3}$ & \cite{rhoelec,rhomassformula}  \\
  \  E/A ($\rho_{sat}$)  & -16.0000 &  0.2 &  MeV &  \cite{rhoelec,rhomassformula} \\
   \ $m^*/m$  & 0.7000  &  0.02 & & \cite{BlaizotKinf,meff} \\
  \  $K_\infty$  &  230.00 & 10.00 & MeV & \cite{BlaizotKinf} \\
   \ $J$   & 32.00 & 2.00 & MeV &   \\
  \hfill 
  \textit{EoS PNM} & & & &\cite{Wiringa} \\
  \ E/N ($\rho$=0.1) & 11.88 & 2.0 & MeV & \\
  \ E/N ($\rho$=0.3) & 35.94 & 7.0 & MeV & \\
   \ E/N ($\rho$=0.35) & 44.14 & 9.0 & MeV & \\ 
\hfill \textit{Stability} & & & & \cite{hel13} \\	
\ INM(S,M,T)& $\rho_{crit} \geq 0.24$ & asymmetric & fm$^{-3}$ &  \\
& &  constraint & &\\
\hline \hline
\ \ \textbf{Finite nuclei } \ \ & & & & \\
  \hfill \textit{Binding energies} & & & & \cite{wan12}\\
 \ $^{40}$Ca \hfill & -342.02300  & 1.5  & MeV &  \\
 \  $^{48}$Ca \hfill  & -415.98300 & 1.0  & MeV &  \\
  \ $^{56}$Ni \hfill  & -483.95300 &  1.5  & MeV  &  \\
   \ $^{100}$Sn \hfill &  -825.13000 &  1.5  & MeV & \\
  \ $^{132}$Sn \hfill & -1102.67300 & 1.0  & MeV  & \\
  \ $^{208}$Pb \hfill & -1635.86100 &   1.0 & MeV  & \\
  \hline 
  \hfill \textit{Proton radii } & & & &  \cite{ang04} \\
  \  $  ^{40}$Ca \hfill &  3.38282  & 0.03 & fm &   \\
   \  $ ^{48}$Ca \hfill  & 3.39070 & 0.02 & fm  & \\
  \   $ ^{56}$Ni \hfill  &  3.66189 & 0.03 &  fm  & \\
   \ $^{132}$Sn \hfill &  4.64745 & 0.02 & fm  & \\
  \  $^{208}$Pb \hfill & 5.45007 & 0.02  & fm  & \\
\hline
\hline
\quad \qquad \textbf{Parameter $W_0$} \ \ &  120.0  & 2.0 & MeV.fm$^5$ &   \\
\hline
\end{tabular}
\caption{Constraints $\mathcal{O}_i$ used in the fitting procedure and the associated error $\Delta \mathcal{O}_i$. See text for details.}
\label{Totcont}
\end{table}

At the end of the minimisation procedure, we have obtained the parameters $\mathbf{p}=\left\{ t_0,t_1,t_2,\dots \right\}$ given in Tab.~\ref{tab:inter}. Notice that the exponent $\alpha$ of the density dependent term has been fixed from the beginning (see Sec.~\ref{sec:infm}). From the table, it is difficult to judge the quantitative relative importance of the different parameters. A way to bypass the problem is to use the concept of \emph{naturalness}. Following Ref.~\cite{kor10B} we multiply each N2LO coupling constant by
\begin{equation}
S=f_\pi^{2(l-1)}\Lambda^{n+l-2}\;,
\end{equation}
where $f_\pi=93$ MeV is the pion decay constant, $\Lambda=687$ MeV, $l$ is the power of the density of the corresponding term and $n$ is the order.
Special treatment is required for the density dependent coupling constant. See Ref.~\cite{kor10B} for details. It is important to keep in mind that the value of $\Lambda$ is somehow arbitrary since it has been derived in Ref.~\cite{kor10B} by observing the behaviour of several N1LO functionals. The results are presented in Tab.~\ref{tab:natural}. Owing to the arbitrariness of the value of $\Lambda$, one should not look too close to the actual numbers, but only to the order of magnitude. By inspecting the table, we clearly observe that there is a natural hierarchy in the coupling constants: the N2LO coupling constants are one order of magnitude smaller than the N1LO ones. This is a very important aspect since the entire idea behind the N$\ell$LO expansion is to have a fast convergence: from these results, we can expect that within this scheme, the N3LO coupling constants would be another order of magnitude smaller.
\begin{table}
\begin{center}
\begin{tabular}{cc|cc}
\hline
\hline
 \multicolumn{4}{c}{SN2LO1}\\
\hline 
$n$ & $i$ & $t_i^{(n)}$ [MeVfm$^{3+n}$] & $x_i^{(n)}$\\
\hline
 0 & 0 & -2486.90786& 0.57160369\\
 2 & 1 &  497.51821 &   -0.05521333\\
 2 & 2&   -451.60715 &   -0.99803779\\
4 & 1 &  -11.95063&    0.10279808\\
4 & 2 &  -15.04405 & -0.93024200\\
\hline
\multicolumn{4}{c}{$t_3=13707.18320$ [MeVfm$^{3(1+\alpha)}$]  $x_3=0.88704830$}\\
\hline
\multicolumn{4}{c}{$\alpha$  = 1/6}\\
\multicolumn{4}{c}{$W_0$  = 117.904418   [MeVfm$^5$]}\\
\hline
\hline
\end{tabular}
\end{center}
\caption{Numerical values for N2LO parameters. }
\label{tab:inter}
\end{table}

\begin{table}
\begin{center}
\begin{tabular}{cc|cc}
\hline
\hline
 \multicolumn{4}{c}{SN2LO1}\\
  \multicolumn{4}{c}{Natural units}\\
\hline 
 $C^{\rho}_0$ & -1.06 &  $C^{\rho}_1$ & 0.754 \\
$C^{\rho}_0 [\rho^\alpha]$ &13.0 &  $C^{\rho}_1 [\rho^\alpha]$ & -12.1 \\
  $C^{\tau}_0$ & 0.892 &  $C^{\tau}_1$ & 0.00624 \\
   $C^{\Delta\rho}_0$ & -1.06 &  $C^{\Delta\rho}_1$ & 0.382 \\
    $C^{\nabla J}_0$ & -1.22  &  $C^{\nabla J}_1$ & -0.406 \\
     $C^{T}_0$ & -0.0882 &  $C^{T}_1$ & -0.816 \\
     \hline
     $C^{(\Delta\rho)^2}_0$ & -0.115  &  $C^{(\Delta\rho)^2}_1$ & 0.0396  \\ 
      $C^{M \rho}_0$ & -0.288 &  $C^{M \rho}_1$ & 0.143 \\
      $C^{M s}_0$ & 0.117  &  $C^{M s}_1$ & -0.0162  \\
\hline
\hline
\end{tabular}
\end{center}
\caption{Values of the parameters of the N2LO pseudo-potential expressed in natural units.}
\label{tab:natural}
\end{table}

\subsection{Finite-size instabilities}\label{sec:finitesize}

As discussed in the introduction, several effective interactions are biased by spurious instabilities~\cite{mar02,report,dep16}. To avoid such a problem, we have developed in Ref.~\cite{pas13} a new fitting protocol based on the LR formalism~\cite{bec14}. From previous analysis of Refs~\cite{hel13,Pas15T}, we have noticed that when a pole in the response function appears at densities lower than $\approx1.2$ saturation density then it is very likely to observe an instability also in the atomic nucleus. Of course, such a criterion does not apply to the spinodal instability, that has a well-defined physical meaning~\cite{duc07}. We have thus added such an additional constraint on top of our fitting protocol to guarantee stable results (see Eq.~\ref{eq:chi2:spurious}). In principle, finite-size instabilities may appear in isospin asymmetric matter as well, see discussion in Ref.~\cite{dav14a}. However, we have not derived the LR formalism for the N2LO functional in this case: as an empirical rule, we decided to add a check on the behaviour of finite size-instabilities also in pure neutron matter even if this does not guarantee that an instability may appear at lower critical density for some specific asymmetry value. At present, such a check is not possible and we leave this aspect for a near future investigation. 

We start by considering the properties of Landau parameters~\cite{lan59}. Their calculation for an extended Skyrme pseudo-potential has been reported in Ref.~\cite{dav14c}. These parameters can be related to properties of infinite nuclear medium and help us constraining some important parts of the effective interaction~\cite{dav14c,Dav16H,back75,zuo03}. In Fig.~\ref{landau}, we show the density dependence of the Landau parameters in SNM. We observe that apart from the physical spinodal instability observed in the $F_0$ parameters, all the Landau inequalities~\cite{mar02} are respected up to two times the saturation density. The only instability appears in the $G'_0$ parameter at $\rho\approx0.35$ fm$^{-3}$. This does not represent a major issue for this study since we do consider only finite-nuclei and and not astrophysical applications~\cite{gor10}.
\begin{figure}[!h]
\begin{center}
\includegraphics[width=0.45\textwidth,angle=-90]{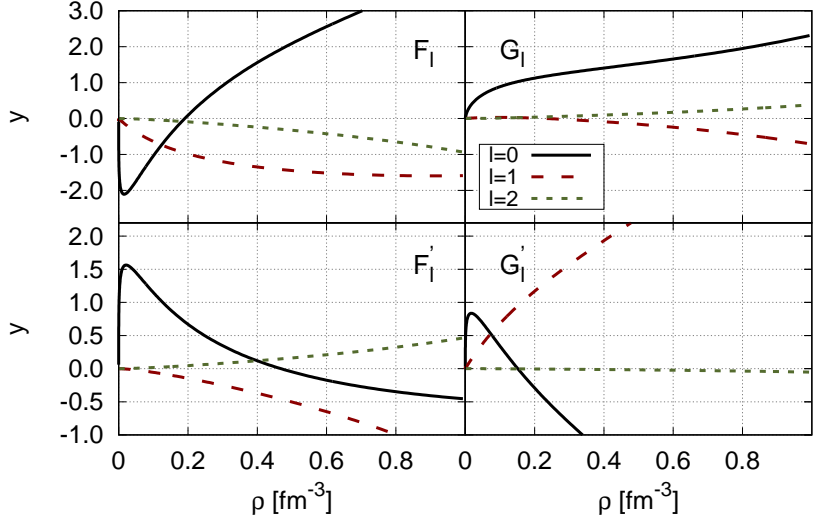}
\end{center}
\caption{(Colors online) Landau parameters in SNM for the SN2LO1 pseudo-potential as a function of the density of the system. See text for details. }
\label{landau}
\end{figure}
In Fig.~\ref{lr}, we show the position of the critical densities obtained in SNM as a function of the transferred momentum $q$. The LR is calculated for each spin (S) spin projection (M)  and isospin (I) channel (S,M,I). See Ref~\cite{report} for more details on the adopted notation. We observe no finite-size instabilities, apart from the physical spinodal one~\cite{duc07}, around saturation density. This means that our interaction is well stable in all spin-isospin channels~\cite{hel13,Pas15T}.  This results confirm our preliminary findings in Ref.~\cite{pas13}: the LR formalism can be considered as a very simple tool to be added in a fitting procedure to avoid exploring regions of parameters that induce unphysical instabilities.
\begin{figure}[!h]
\begin{center}
\includegraphics[width=0.45\textwidth]{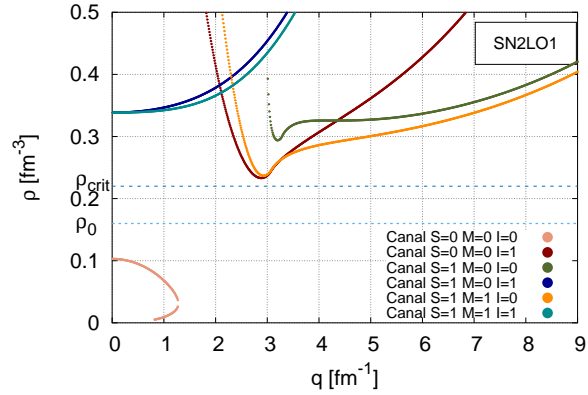}
\end{center}
\caption{(Colors online) Critical densities in SNM as a function of transferred momentum $q$. The horizontal dashed lines represent the saturation density $\rho_0$ and the critical density $\rho_{crit}$.}
\label{lr}
\end{figure}

\begin{figure}[!h]
\begin{center}
\includegraphics[width=0.50\textwidth,angle=-90]{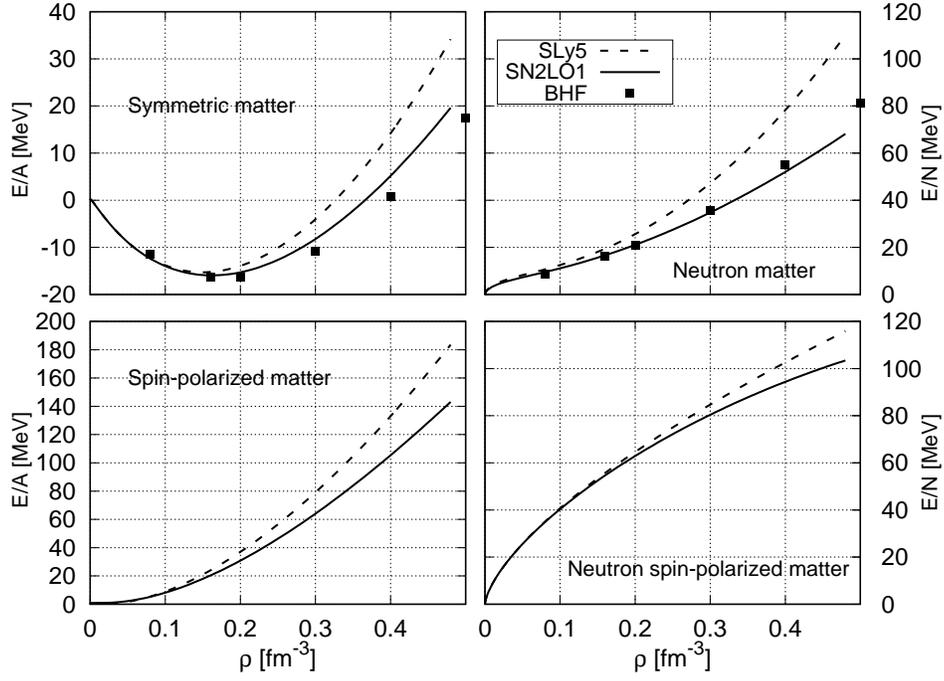}
\end{center}
\caption{(Colors online) Equation of state for SNM and PNM obtained with the N2LO Skyrme interaction. The squares represent the values obtained from BHF calculations. }
\label{eos:tot}
\end{figure}

\subsection{Infinite nuclear matter}\label{sec:infm}

In our fitting protocol, we include information of the infinite nuclear medium. Following Ref.~\cite{cha97}, we have used as a constrain three points of the  EoS in PNM dervied in Ref~\cite{Wiringa}. We can now benchmark our results against other well known EoS as the one derived via Brueckner-Hartree-Fock (BHF)~\cite{bal97}.
In Fig.~\ref{eos:tot}, we compare the EoS for symmetric matter and neutron matter obtained with BHF and the SN2LO1 interaction. For completeness the results with SLy5 are also given. The SN2LO1 follows quite closely the BHF results, and in particular the EoS of PNM up to 3 times saturation density. Beyond this point the EoS becomes slightly softer. We remind the reader that SLy5 and SN2LO1 follow each other quite closely in PNM at low-density since they have been constrained on the same points in this density region.

On the same figure, we also give the results for spin-polarised symmetric matter and spin-polarised pure neutron matter and compare SLy5 and SN2LO1 results. Although these two quantities have not been fitted explicitly, we observe a qualitative similar behaviour in the two functionals. For completeness, in Tab.~\ref{tab:inm}, we give the main features of the EoS of SN2LO1, $i.e.$ saturation density $\rho_0$, incompressibility $K_{\infty}$, symmetry energy $J$ and slope of symmetry energy $L$ (not fitted). The values we obtained are in agreement with the existing constraints~\cite{dut12}. 
 
 As already discussed in Ref.~\cite{cha97}, there is a strong model correlation for N1LO between the nuclear incompressibility and the effective mass. In our case, the correlation between  $K_\infty$ and ${m}/{ m^*}$ is of course different since the new parameters give us more freedom in adjusting these two values. It can be calculated analytically in infinite matter with the result
\begin{equation}\label{corrN2LO}
K_\infty = - 9 (\alpha+1) \frac{E}{A}(\rho_0) + \frac{3}{5} \frac{\hbar^2}{2m} k_F^2 \left( 3 \ (3 \alpha-1) - 2 \ (3 \alpha - 2)  \frac{m}{\ m^*} \right) + \frac{3}{140} C_0^{(4)} \rho k_F^4 ( 3 \alpha + 10)\;. 
\end{equation}
In Fig.~\ref{corr:kinf}, we observe that to obtain a reasonable value of the nuclear incompressibility, the allowed range for $\alpha$ is $\alpha\in[1/6,1/3]$. In a future work, we plan to remove the density dependent term and to replace it with a real three-body term~\cite{sad13} to make the pseudo-potential suitable also for multi-reference calculations~\cite{lac09,ben09}.
\begin{table}
\begin{center}
\begin{tabular}{c|c|c}
\hline
 & SN2LO1 & SLy5 \\
\hline
$\rho_0$ [fm$^{-3}$] &0.162 &0.1603 \\
E/A($\rho_0$)  [MeV]& -15.948&-15.98\\
$K_{\infty}$  [MeV] &221.9 & 229.92\\
$J$   [MeV] & 31.95 & 32.03\\
$L$  [MeV] & 48.9 &48.15\\
$m^*/m$ & 0.709 & 0.696 \\
\hline
\hline
\end{tabular}
\end{center}
\caption{Infinite matter properties at saturation for SN2LO1 and SLy5~\cite{cha97}. See text for details. }
\label{tab:inm}
\end{table}

\begin{figure}[!h]
\begin{center}
\includegraphics[width=0.45\textwidth,angle=0]{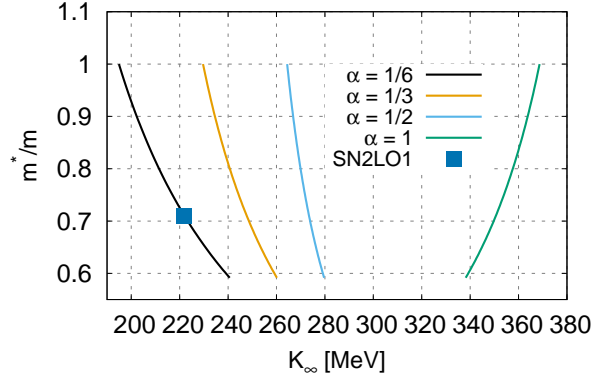}
\end{center}
\caption{(Colors online) Correlation between the effective mass and the nuclear incompressibility in infinite nuclear matter for different values of the power of the density dependent term.}
\label{corr:kinf}
\end{figure}

\subsection{Finite nuclei}

In this section, we analyse the properties of finite nuclei obtained with the extended Skyrme pseudo-potential. In Fig.~\ref{tikzmass}, we show the energy difference $\Delta E$ between the experimental values and the ones calculated using either SLy5 or SN2LO1 for the few selected double-magic nuclei used in the fit. The results obtained with  SN2LO1 are of the same quality as SLy5. Moreover they are all very close to the tolerance $\Delta \mathcal{O}_i$ we used for the fit given in Tab.~\ref{Totcont}.
\begin{figure}
\centering
\begin{tikzpicture}[scale=0.6]
\draw (-5,0) grid (5,7);
\draw (0,-0.7) node[below]{\large $\Delta$E =  E$_{\mbox{\small{th}}}$ - E$_{\mbox{\small{exp}}}$ \quad [MeV]};
\draw[line width = 1.5pt] (0,0) -- (0,7);
\draw(-5.2,1)node[left]{\Large $^{40}$ Ca};
\draw(-5.2,2)node[left]{\Large $^{48}$ Ca};
\draw(-5.2,3)node[left]{\Large $^{56}$ Ni};
\draw(-5.2,4)node[left]{\Large $^{100}$ Sn};
\draw(-5.2,5)node[left]{\Large $^{132}$ Sn};
\draw(-5.2,6)node[left]{\Large $^{208}$ Pb};
\foreach \x in {-5,...,5} \draw(\x,0)node[below]{\x};
\draw[white, fill, rounded corners,opacity=0.9] (1.4,0.6) rectangle (4.9,2.4);
\draw (3.2,2) node {\Large $\bullet$ SN2LO1};
\draw (2.6,1) node[color=orange] {\Large $\bullet$ SLy5};
\draw[line width = 4pt] plot[xcomb,mark=*] coordinates
{(-345.36828+342.02300,1.15) (-416.94469+415.98300,2.15)(-482.12111+483.95300,3.15)(-826.06845+825.13000,4.15)(-1104.44457+1102.67300,5.15)(-1634.36724+1635.86100,6.15)};
\draw[line width = 4pt, color=orange] plot[xcomb,mark=*] coordinates
{(-344.05099+342.02300,0.85)(-415.90143+415.98300,1.85)(-482.65365+483.95300,2.85)(-827.80765+825.13000,3.85)(-1103.86452+1102.67300,4.85)(-1635.97989+1635.86100,5.85)};
\end{tikzpicture}
\caption{Difference of binding energies obtained with SN2LO1 and SLy5 and experimental values extracted from Ref. \cite{wan12}.}
\label{tikzmass}
\end{figure}
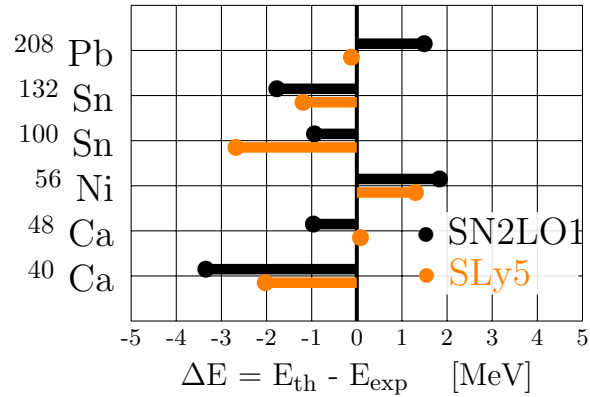
In Fig.~\ref{tikzrad}, we compare the differences of proton radii $\Delta r_p$ obtained with SLy5 and our new pseudo-potential SN2LO1. In this case we see that SN2LO1 behaves marginally better than SLy5 giving a result typically closer to the experimental values.
 It is worth noticing that compared to SLy5, we have few additional constraints concerning finite-size instabilities that were not present in the original fitting protocol of SLy5. The closest functional to SN2LO1, in terms of fitting protocol, is represented by SLy5$^*$~\cite{pas13}. We do not report here the direct comparison, but we have checked that the results are qualitatively the same.
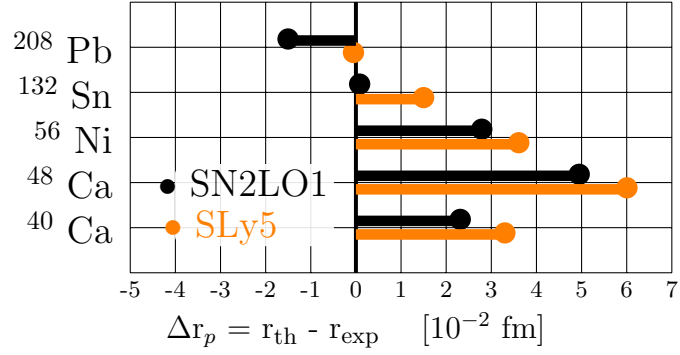
\begin{figure}
\centering
\begin{tikzpicture}[scale=0.6]
\draw (-5,0) grid (7,6);
\draw (0,-0.7) node[below]{\large $\Delta$r$_p$ =  r$_{\mbox{\small{th}}}$ - r$_{\mbox{\small{exp}}}$ \quad [10$^{-2}$ fm]};
\draw[line width = 1.5pt] (0,0) -- (0,6);
\draw(-5.2,1)node[left]{\Large $^{40}$ Ca};
\draw(-5.2,2)node[left]{\Large $^{48}$ Ca};
\draw(-5.2,3)node[left]{\Large $^{56}$ Ni};
\draw(-5.2,4)node[left]{\Large $^{132}$ Sn};
\draw(-5.2,5)node[left]{\Large $^{208}$ Pb};
\draw (340.594-338.282,1.15) node {\huge $\bullet$};
\draw (344.018-339.070,2.15) node {\huge $\bullet$};
\draw (368.978-366.189,3.15) node {\huge $\bullet$};
\draw (464.828-464.745,4.15) node {\huge $\bullet$};
\draw (543.504-545.007,5.15) node {\huge $\bullet$};
\draw (341.583-338.282,0.85) node[color=orange] {\huge $\bullet$};
\draw (345.076-339.070,1.85) node[color=orange] {\huge $\bullet$};
\draw (369.798-366.189,2.85) node[color=orange] {\huge $\bullet$};
\draw (466.245-464.745,3.85) node[color=orange] {\huge $\bullet$};
\draw (544.953-545.007,4.85) node[color=orange] {\huge $\bullet$};
\draw[white, fill, rounded corners,opacity=0.9] (-4,0.6) rectangle (-0.5,2.4);
\draw (-2.5,2) node {\Large $\bullet$ SN2LO1};
\draw (-3,1) node[color=orange] {\Large $\bullet$ SLy5};
\foreach \x in {-5,...,7} \draw(\x,0)node[below]{\x};
\draw[line width = 4pt,xscale=100.0,yscale=1.0] plot[xcomb] coordinates
{(3.40594-3.38282,1.15)(3.44018-3.39070,2.15)(3.68978-3.66189,3.15)(4.64828-4.64745,4.15)(5.43504-5.45007,5.15)};
\draw[line width = 4pt,xscale=100.0,yscale=1.0,color=orange] plot[xcomb] coordinates
{(3.41583-3.38282,0.85)(3.45076-3.39070,1.85)(3.69798-3.66189,2.85)(4.66245-4.64745,3.85)(5.44953-5.45007,4.85)};
\end{tikzpicture}
\caption{Proton radii difference of two interactions (SN2LO1/SLy5) calculated with WHISKY with experimental radii obtained in Ref.~\cite{ang04}.}
\label{tikzrad}
\end{figure}

In Figs.\ref{be}, we compare the differences between the binding energies calculated for isotopic (isotonic) chains with Z(N)=20, 28, 50, 82 for our extended Skyrme interaction. The experimental measurements are taken from Ref.~\cite{wan12}. On the same figure, we also report the values obtained with SLy5. Notice that we did not optimised the value of the pairing strength to improve the reproduction of experimental data. Moreover, since the effective masses are numerically quite similar for SLy5 and SN2LO1, we used exactly the same pairing interaction.
\begin{figure}[!h]
\begin{center}
\includegraphics[width=0.34\textwidth,angle=-90]{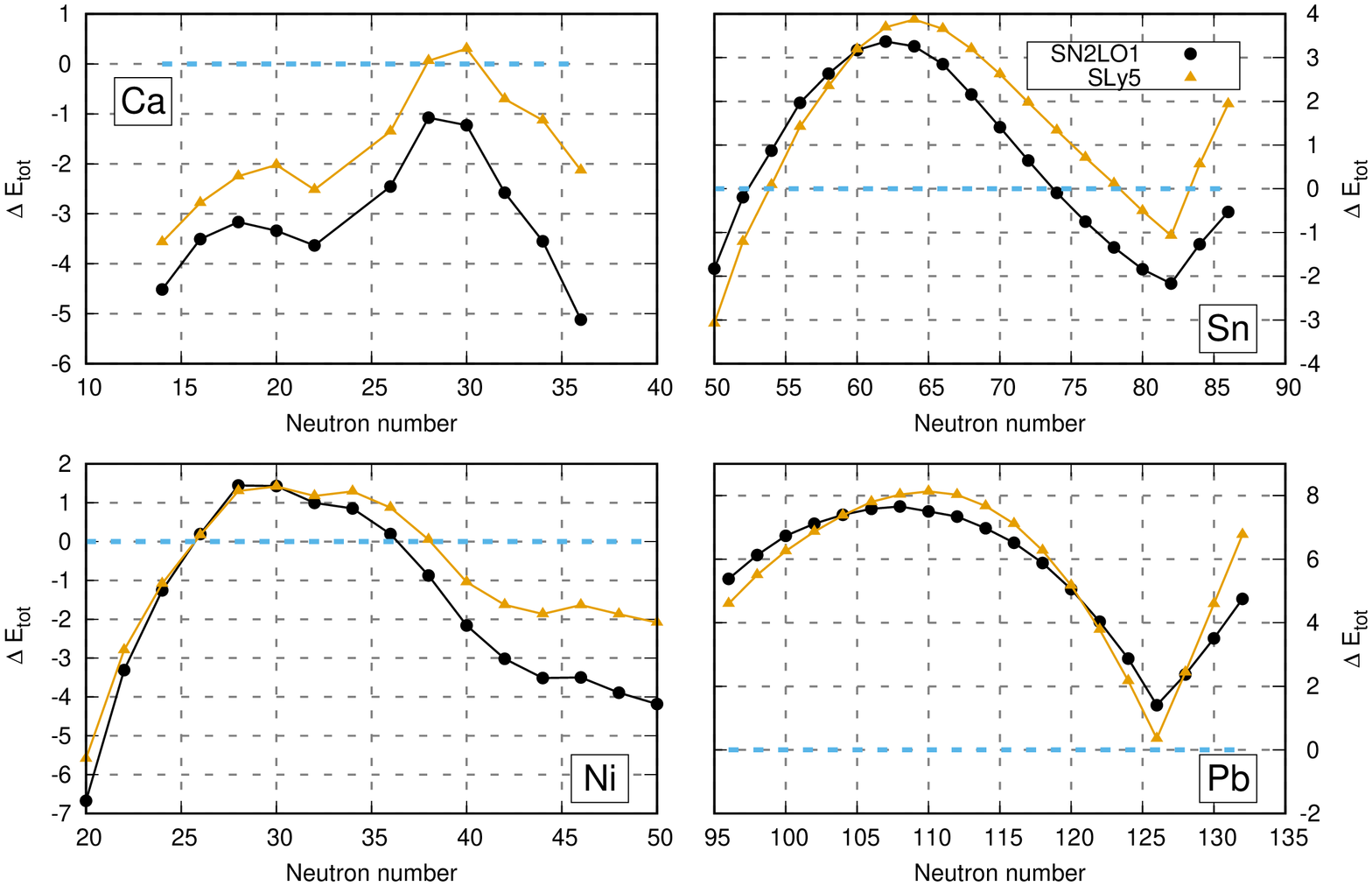}
\hspace{-0.3cm}
\includegraphics[width=0.34\textwidth,angle=-90]{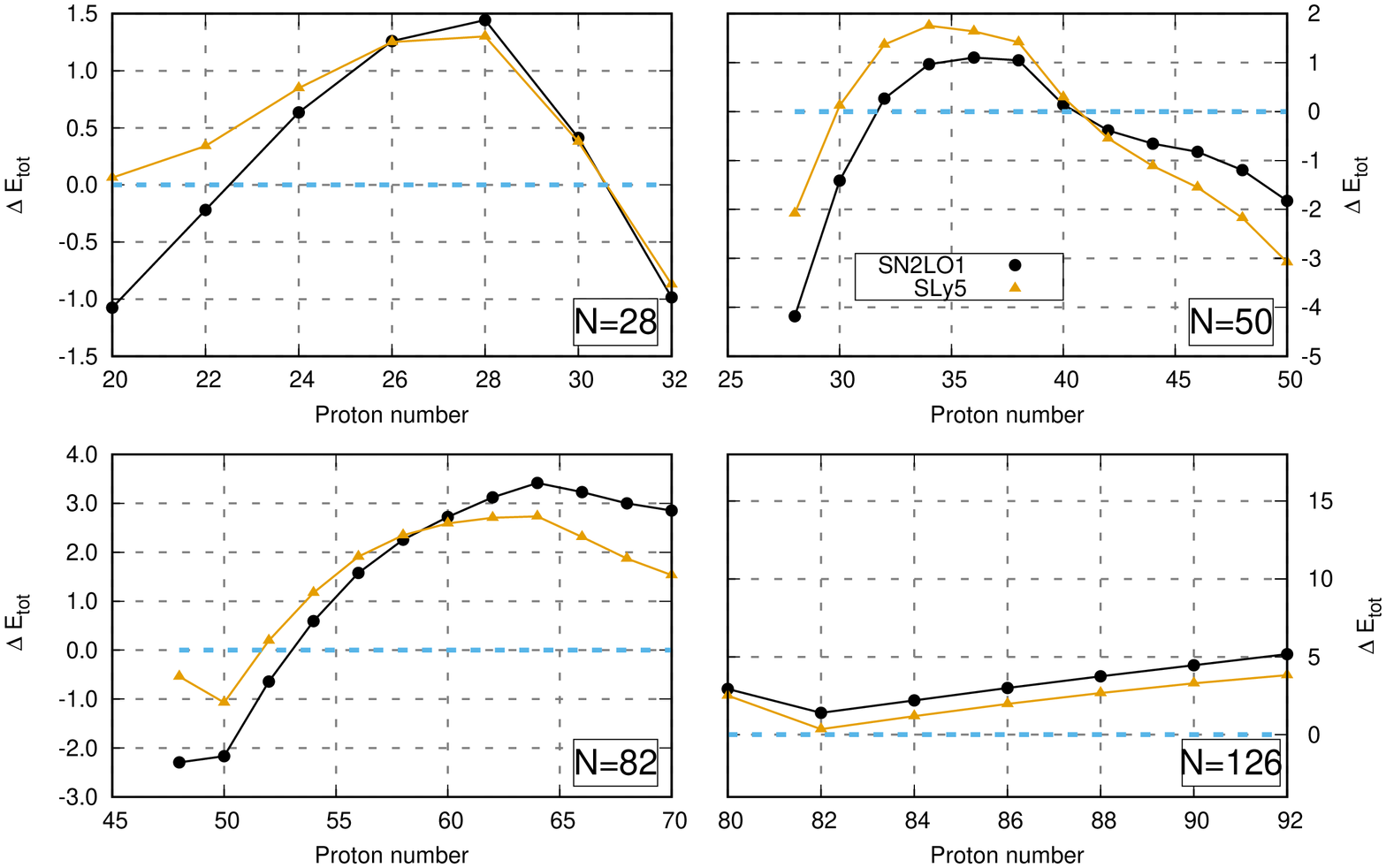}
\end{center}
\caption{(Colors online) Systematic comparison of binding energies, expressed in MeV, for isotopic (isotonic) chains calculated with our extended Skyrme  interaction SN2LO1 and experimental ones. On the same figure we also compare with the SLy5 parametrisation. See text for details.}
\label{be}
\end{figure}
The main feature we observe is the strong arch-like structures. This is the main drawback of a fitting protocol that fixes a very limited number of nuclei. A better fitting protocol has been designed for example  for UNEDF functionals~\cite{kor10,kor12,kor13} and we plan to use it for a systematic exploration of the parameter space of higher order terms. In Fig.~\ref{charge}, we compare the proton radii. The data are taken from Ref.~\cite{ang04}. The new interaction is fairly closer to experimental data than the original SLy5 and the main trends are reproduced. One of the biggest discrepancy we observe in the data is related to the anomalous isotopic dependence of proton radii of calcium isotopes. With the current parametrisation we have not been able to reproduce both $^{40}$Ca and $^{48}$Ca.
A recent article~\cite{rei17} suggests that a different form of the pairing functionals based on Fayans form~\cite{fay00} may be the key to solve this anomaly, while the specific form of the functional used for the calculation of the central potential is not relevant. Since we did not fix any particular pairing functional in our fit, we plan to test the results of Ref.~\cite{rei17} with our new functional.

Finally, we have explored the behaviour of single particle spectra. In Fig.~\ref{fig:sparticle40ca}, we compare the Hartree-Fock neutron single particle states for $^{40}$Ca obtained using SLy5 and SN2LO1. The values are compared with the experimental values extracted from Ref.~\cite{Sch07}. The HF states obtained with the two functionals are very close to each other. SN2LO1 shows a slight compression of the spectrum, but this is simply related to a slightly larger effective mass (see Tab.~\ref{tab:inm}). Similar behaviour is also observed in Fig.~\ref{fig:sparticle208pb} for neutron single-particle states in $^{208}$Pb.

As discussed in Sec.~\ref{sec:hfb}, the higher order gradient terms induce three extra spin-orbit fields Eqs.\ref{eq:so1}-\ref{eq:so4}. In principle this should provide us with some extra flexibility compared to a standard Skyrme interaction. However, the major problem encountered in this first analysis is to find the right observable that may let us explore a new region of parameter space that may increase their importance. We recall that we neglected completely tensor terms at N2LO level, which means two extra tensor parameters~\cite{Dav15,Dav16AN}. This could also give an extra freedom to correct some known anomaly in the shell evolution of some particular states~\cite{col07}. The exploration of this particular aspect is currently under investigation.
\begin{figure}[!h]
\begin{center}
\includegraphics[width=0.4\textwidth,angle=-90]{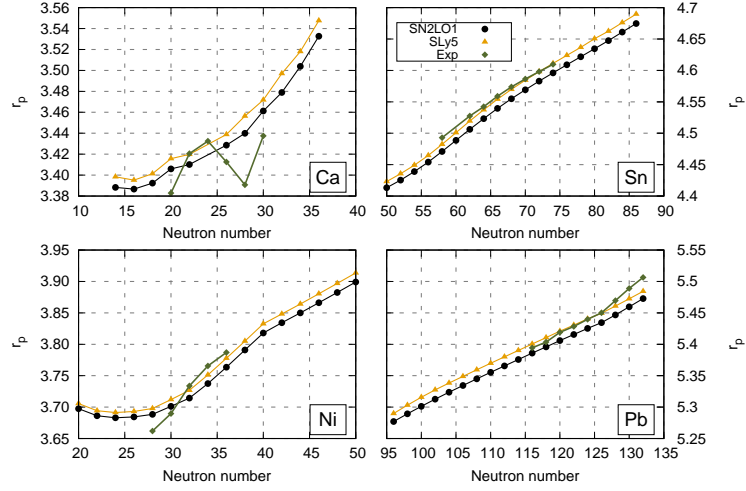}
\end{center}
\caption{(Colors online) Systematic comparison of proton radii. Experimental data are taken from Ref.~\cite{ang04}.}
\label{charge}
\end{figure}

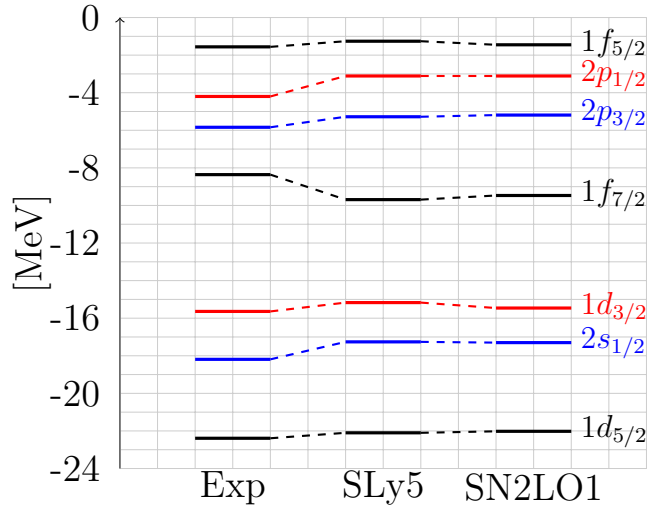
\begin{figure}
\centering
\begin{tikzpicture}[scale=0.5]
\draw [->, yscale = 0.5] (0,-24) -- (0,0);
\draw [very thin, color=gray, opacity = 0.4, yscale=0.5] (0,-24) grid[step=1] (14		,0);
\draw (-1.6,-6) node[above,rotate=90] {\Large  [MeV]};
\foreach \y in {-24, -20,...,0} \draw(-0.3,\y*0.5)node[left]{\Large \y};
\draw[yscale = 0.5] ( 3,-25) node {\Large Exp};
\black\draw [very thick, yscale = 0.5] (2,-22.39) -- (4,-22.39);
\blue\draw [very thick, yscale = 0.5] (2,-18.19) -- (4,-18.19) ;
\red\draw [very  thick, yscale = 0.5] (2,-15.64) -- (4,-15.64) ;
\black\draw [very  thick, yscale = 0.5] (2,-8.36) -- (4,-8.36);
\blue\draw [very  thick, yscale = 0.5] (2,-5.84) -- (4,-5.84) ;
\red\draw [very  thick, yscale = 0.5] (2,-4.20) -- (4,-4.20) ;
\black\draw [very thick, yscale = 0.5] (2,-1.56) -- (4,-1.56) ;
\draw[yscale = 0.5] ( 7,-25) node {\Large SLy5};
\black\draw [very  thick, yscale = 0.5] (6,-22.10) -- (8,-22.10) ;
\blue\draw [very  thick, yscale = 0.5] (6,-17.26) -- (8,-17.26) ;
\red\draw [very  thick, yscale = 0.5] (6,-15.17) -- (8,-15.17) ;
\black\draw [very  thick, yscale = 0.5] (6,-9.69) -- (8,-9.69);
\blue\draw [very thick, yscale = 0.5] (6,-5.28) -- (8,-5.28);
\red\draw [ very  thick, yscale = 0.5] (6,-3.11) -- (8,-3.11);
\black\draw [very  thick, yscale = 0.5] (6,-1.26) -- (8,-1.26);
\draw[yscale = 0.5] ( 11,-25) node {\Large SN2LO1};
\black\draw [very  thick, yscale = 0.5] (10,-22.02) -- (12,-22.02) node[right]{\large 1$d_{5/2}$};
\blue\draw [very  thick, yscale = 0.5] (10,-17.30) -- (12,-17.30) node[right]{\large 2$s_{1/2}$};
\red\draw [very  thick, yscale = 0.5] (10,-15.46) -- (12,-15.46) node[right]{\large 1$d_{3/2}$};
\black\draw [very  thick, yscale = 0.5] (10,-9.47) -- (12,-9.47) node[right]{\large 1$f_{7/2}$};
\blue\draw [very  thick, yscale = 0.5] (10,-5.19) -- (12,-5.19) node[right]{\large 2$p_{3/2}$};
\red\draw [very  thick, yscale = 0.5] (10,-3.11) -- (12,-3.11) node[right]{\large 2$p_{1/2}$};
\black\draw [very  thick, yscale = 0.5] (10,-1.45) -- (12,-1.45) node[right]{\large 1$f_{5/2}$};
\black\draw [dashed,thick, yscale = 0.5] (4,-22.39) -- (6,-22.10) ;
\black\draw [dashed,thick, yscale = 0.5] (8,-22.10)  -- (10,-22.02) ;
\blue\draw [dashed,thick, yscale = 0.5] (4,-18.19)--(6,-17.26) ;
\blue\draw [dashed,thick, yscale = 0.5] (8,-17.26)  -- (10,-17.30)  ;
\red\draw [dashed,thick, yscale = 0.5] (4,-15.64)--(6,-15.17)  ;
\red\draw [dashed,thick, yscale = 0.5] (8,-15.17)  --  (10,-15.46)  ;
\black\draw [dashed,thick, yscale = 0.5](4,-8.36)--(6,-9.69)  ;
\black\draw [dashed,thick, yscale = 0.5] (8,-9.69)  -- (10,-9.47) ;
\blue\draw [dashed,thick, yscale = 0.5](4,-5.84)--(6,-5.28)   ;
\blue\draw [dashed,thick, yscale = 0.5](8,-5.28)  -- (10,-5.19) ;
\red\draw [dashed,thick, yscale = 0.5](4,-4.20)--(6,-3.11)    ;
\red\draw [dashed,thick, yscale = 0.5](8,-3.11)   -- (10,-3.11) ;
\black\draw [dashed,thick, yscale = 0.5](4,-1.56)--(6,-1.26)   ;
\black\draw [dashed,thick, yscale = 0.5] (8,-1.26)   --(10,-1.45)  ;
\end{tikzpicture}
\caption{Neutron single-particle energies around the Fermi energy in the $^{40}$Ca  for SLy5 and SN2LO1 parametrisations. The experimental values are taken from Ref.~\cite{Sch07}. See text for details.}
\label{fig:sparticle40ca}
\end{figure}

\begin{figure}
\centering
\begin{tikzpicture}[scale=0.5]
\draw [->] (0,-13) -- (0,0);
\draw [very thin, color=gray, opacity = 0.4] (0,-13) grid[step=1] (14,0);
\draw (-1.4,-7) node[above,rotate=90] { \Large [MeV]};
\foreach \y in { -12,-10,...,0} \draw(-0.3,\y)node[left]{\Large \y};
\draw ( 3,-13) node[below] {\Large Exp};
\black\draw [very thick] (2,-11.40) -- (4,-11.40);
\blue\draw [very thick] (2,-9.81) -- (4,-9.81) ;
\red\draw [very  thick] (2,-9.24) -- (4,-9.24) ;
\black\draw [very  thick] (2,-8.26) -- (4,-8.26);
\blue\draw [very  thick] (2,-7.94) -- (4,-7.94) ;
\red\draw [very  thick] (2,-7.37) -- (4,-7.37) ;
\black\draw [very thick] (2,-3.94) -- (4,-3.94) ;
\blue\draw [very thick] (2,-3.16) -- (4,-3.16) ;
\black\draw ( 7,-13) node[below] {\Large SLy5};
\black\draw [very  thick] (6,-12.76) -- (8,-12.76) ;
\blue\draw [very  thick] (6,-12.09) -- (8,-12.09) ;
\red\draw [very  thick] (6,-9.40) -- (8,-9.40) ;
\black\draw [very  thick] (6,-9.25) -- (8,-9.25);
\blue\draw [very thick] (6,-9.13) -- (8,-9.13);
\red\draw [ very  thick] (6,-8.15) -- (8,-8.15);
\black\draw [very  thick] (6,-3.2) -- (8,-3.2);
\blue\draw [very  thick] (6,-1.91) -- (8,-1.91);
\black\draw  ( 11,-13) node[below] {\Large SN2LO1};
\black\draw [very  thick] (10,-12.69) -- (12,-12.69) node[right]{1$h_{9/2}$};
\blue\draw [very  thick] (10,-12.11) -- (12,-12.11) node[right]{2$f_{7/2}$};
\red\draw [very  thick] (10,-9.47) -- (12,-9.47) node[below right]{1$i_{13/2}$};
\black\draw [very  thick] (10,-9.37) -- (12,-9.37) node[right]{3$p_{3/2}$};
\blue\draw [very  thick] (10,-9.27) -- (12,-9.27) node[above right]{2$f_{5/2}$};
\red\draw [very  thick] (10,-8.31) -- (12,-8.31) node[above right]{3$p_{1/2}$};
\black\draw [very  thick] (10,-3.35) -- (12,-3.35) node[right]{2$g_{9/2}$};
\blue\draw [very  thick] (10,-2.03) -- (12,-2.03) node[right]{1$i_{11/2}$};
\black\draw [dashed,thick] (4,-11.40) -- (6,-12.76) ;
\black\draw [dashed,thick] (8,-12.76)  -- (10,-12.69) ;
\blue\draw [dashed,thick] (4,-9.81)--(6,-12.09) ;
\blue\draw [dashed,thick] (8,-12.09)  -- (10,-12.11)  ;
\red\draw [dashed,thick] (4,-9.24)--(6,-9.40)  ;
\red\draw [dashed,thick] (8,-9.40)  --  (10,-9.47)  ;
\black\draw [dashed,thick](4,-8.26)--(6,-9.25)  ;
\black\draw [dashed,thick] (8,-9.25)  -- (10,-9.37) ;
\blue\draw [dashed,thick](4,-7.94)--(6,-9.13)   ;
\blue\draw [dashed,thick](8,-9.13)  -- (10,-9.27) ;
\red\draw [dashed,thick](4,-7.37)--(6,-8.15)    ;
\red\draw [dashed,thick](8,-8.15)   -- (10,-8.31) ;
\black\draw [dashed,thick](4,-3.94)--(6,-3.2)   ;
\black\draw [dashed,thick] (8,-3.2)   --(10,-3.35)  ;
\blue\draw [dashed,thick](4,-3.16)--(6,-1.91)   ;
\blue\draw [dashed,thick] (8,-1.91)   --(10,-2.03)  ;
\end{tikzpicture}
\caption{Same as Fig.~\ref{fig:sparticle40ca}, but for $^{208}$Pb. }
\label{fig:sparticle208pb}
\end{figure}
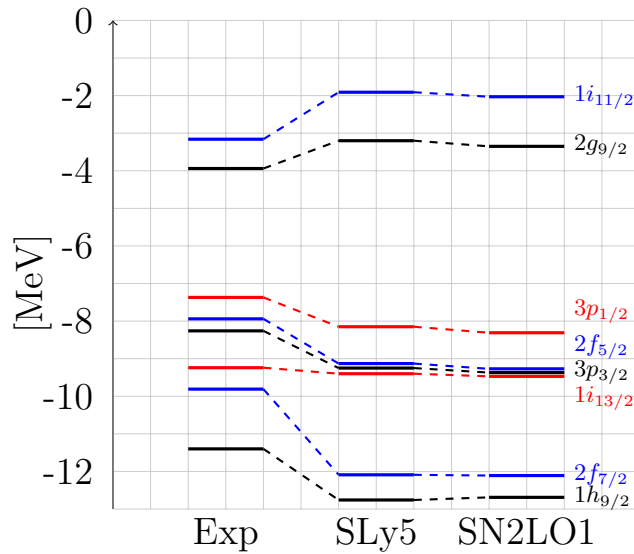

\section{Conclusions}\label{sec:conclusions}

In the present article, we have discussed the formalism to include fourth-order gradient terms of the N2LO Skyrme interaction. We have derived the functional, the complete expression of the densities in the case of spherical symmetry and the corresponding HF equation. The resulting 4-th order differential equation has been solved with a new numerical code named \verb|WHISKY|. This code has been tested against two different HFB solvers to check numerical accuracy of the new solver. Thanks to this new code, we have been able to perform for the very first time a complete fit of a stable N2LO Skyrme interaction including finite-nuclei. This achievement has been made possible by the use of the Linear Response formalism as a tool to prevent unphysical instabilities. 

For the very first time, we thus have been able to prove that it is possible to go \emph{beyond} the standard Skyrme interaction by including physically motivated terms. Thanks to the work on the foundations of various non-relativistic effective interactions~\cite{Dav16AN}, we have been able to clarify the inner nature of the higher order gradient terms in the extended N$\ell$LO Skyrme pseudo-potential. The LR formalism we have been able to solve also the long-standing problem of finite-size instabilities in effective functionals. Finite-size instabilities seem to appear in various functionals not only the Skyrme-like ones~\cite{dep16}. The LR formalism thus represents a simple tool that should be included in all modern fitting protocol to avoid the appearance of non physical results.

Combining all the previous results, we have been able to derive the complete set of parameters of the N2LO pseudo-potential, named SN2LO1 in this paper. We have compared its performances on both infinite nuclear matter (pseudo)-observables as well as ground state properties of some selected nuclei. The global performances are of the same quality as the standard SLy5. However, it is very important to underline here that since SN2LO1 has four additional parameters compared to SLy5, we have imposed extra stability constraints to our functional: SLy5 has a finite-size instability in the spin-channel and thus can not be used to perform calculations where the time-odd channel is open. To the best of our knowledge, SN2LO1 is free from pathologies and it can be safely used in various numerical codes.

Finally we insist on the fact that the higher order terms introduce several new features as for example three new spin-orbit fields that have not been completely investigated in this article and may give rise to new properties of the functional: N2LO clearly offers some new degrees of freedom and goes beyond N1LO.

\section*{Acknowledgments}

We are grateful to W. Ryssens for providing us with the \verb|MOCCA| results, as well to K. Bennaceur for providing us with the \verb|LENTEUR| code as well for fruitful discussion. We also acknowledge interesting discussions with M. Bender. The work of J.N. has been supported by grant FIS2014-51948-C2-1-P, Mineco (Spain).

\begin{appendix}
\section{Coupling constants}\label{app:cc}

In this section we give the explicit expressions of the new coupling constants of N2LO functional in terms of Skyrme parameters. The expression of the coupling constants for the standard Skyrme functional can be found in Ref.~\cite{les07}.
\begin{eqnarray}
\label{eq:taxa}
C_0^{(\Delta \rho)^2} & = &  \tfrac{1}{128}
                         \, \left[ 9 t_1^{(4)} - t_2^{(4)} \left( 5 + 4 x_2^{(4)} \right) \right] \q   \\
C_1^{( \Delta \rho)^2} & = &    - \tfrac{1}{128}
                         \, \left[ 3 t_1^{(4)} \left( 1 + 2 x_1^{(4)} \right)
                                   + t_2^{(4)} \left( 1 + 2 x_2^{(4)} \right) \right]             \q    \\
C_0^{ M \rho}      & = &  \tfrac{1}{32}
                         \, \left[ 3 t_1^{(4)} + t_2^{(4)} \left( 5 + 4 x_2^{(4)} \right) \right] \q  \\
C_1^{ M \rho}      & = &    - \tfrac{1}{32}
                         \, \left[   t_1^{(4)} \left( 1 + 2 x_1^{(4)} \right)
                                   - t_2^{(4)} \left( 1 + 2 x_2^{(4)} \right) \right]             \q  \\
C_0^{ M s}         & = &    - \tfrac{1}{32}
                         \, \left[   t_1^{(4)} \left( 1 - 2 x_1^{(4)} \right)
                                   - t_2^{(4)} \left( 1 + 2 x_2^{(4)} \right) \right]             \q    \\
C_1^{ M s}         & = &    - \tfrac{1}{32}   \, \left[ t_1^{(4)} - t_2^{(4)}  \right]          \q  
\end{eqnarray}

\section{Densities in Cartesian representation}\label{app:dens}

We define the density matrix in coordinate space as in~\cite{rin80}
\begin{eqnarray}
\rho_q(\mathbf{r}\sigma,\mathbf{r}'\sigma')=\frac{1}{2}\rho_q(\mathbf{r},\mathbf{r}')\delta_{\sigma\sigma'}+\frac{1}{2}\mathbf{s}_q(\mathbf{r},\mathbf{r}')\langle \sigma'| \hat{\sigma}|\sigma\rangle \;,
\end{eqnarray}
where
\begin{eqnarray}
\rho_q(\mathbf{r},\mathbf{r}')&=&\sum_{\sigma}\rho_q(\mathbf{r}\sigma,\mathbf{r}'\sigma')\\
\mathbf{s}_q(\mathbf{r},\mathbf{r}')&=&\sum_{\sigma\sigma'}\rho_q(\mathbf{r}\sigma,\mathbf{r}'\sigma')\langle \sigma'| \hat{\sigma}|\sigma\rangle.
\end{eqnarray}

The Skyrme energy density functional up to 2nd order is composed by  seven local densities whose explict expression can be found, for example, in Ref.~\cite{les07}. The extension to fourth order requires the definition of six additional local densities
\begin{eqnarray}
\tau_{\mu\nu, q}(\mathbf{r})&=&\left. \nabla_\mu\nabla_\nu'\rho_q(\mathbf{r},\mathbf{r}') \right|_{\mathbf{r}=\mathbf{r}'}\\
K_{\mu\nu\kappa, q}(\mathbf{r})&=&\left. \nabla_\mu\nabla_\nu's_{\kappa q}(\mathbf{r},\mathbf{r}') \right|_{\mathbf{r}=\mathbf{r}'}\\
\Pi_{\mu, q}(\mathbf{r})&=&\left. \nabla \cdot \nabla' j_{\mu, q}(\mathbf{r},\mathbf{r}') \right|_{\mathbf{r}=\mathbf{r}'}\\
V_{\mu \nu, q}(\mathbf{r})&=&\left. \nabla \cdot \nabla' J_{\mu\nu, q}(\mathbf{r},\mathbf{r}') \right|_{\mathbf{r}=\mathbf{r}'}\\
Q_q(\mathbf{r})&=&\left. \Delta \Delta' \rho_q(\mathbf{r},\mathbf{r}') \right|_{\mathbf{r}=\mathbf{r}'}\\
S_{\mu, q}(\mathbf{r})&=&\left. \Delta \Delta' s_{\mu, q}(\mathbf{r},\mathbf{r}') \right|_{\mathbf{r}=\mathbf{r}'}
\end{eqnarray}

Similarly to the spin-current pseudo-tensor $J_{\mu\nu, q}(\mathbf{r})$, the density $\tau_{\mu\nu, q}(\mathbf{r})$ can be decomposed into a pseudo-scalar, vector and traceless pseudo-tensor term. For more details we refer to Ref.~\cite{bec14}.

\end{appendix}

\bibliography{biblio}

\end{document}